\documentclass[aps,preprint,floats,nofootinbib]{revtex4}
\usepackage{amsmath}
\usepackage{amssymb}
\usepackage{latexsym}
\usepackage{epsf}
\usepackage{feynarts}
\usepackage{graphicx}

\newlength{\defbaselineskip}
\newcommand{\setlinespacing}[1]%
           {\setlength{\baselineskip}{#1 \defbaselineskip}}

\def\beq{\begin{equation}}

\def\eeq{\end{equation}}

\def\bey{\begin{eqnarray}}

\def\eey{\end{eqnarray}}

\def\lsim{\mathrel{\raise.3ex\hbox{$<$\kern-.75em\lower1ex\hbox{$\sim$}}}}

\def\gsim{\mathrel{\raise.3ex\hbox{$>$\kern-.75em\lower1ex\hbox{$\sim$}}}}

\def\lsim{\mathrel{\raise.3ex\hbox{$<$\kern-.75em\lower1ex\hbox{$\sim$}}}} 
\def\gsim{\mathrel{\raise.3ex\hbox{$>$\kern-.75em\lower1ex\hbox{$\sim$}}}} 
\begin{document}

\preprint{
\hfill
\begin{minipage}[t]{3in}
\begin{flushright}
\vspace{0.0in}
FERMILAB--PUB--06--456--A\\
\end{flushright}
\end{minipage}
}

\hfill$\vcenter{\hbox{}}$

\vskip 0.5cm

\title{Distinguishing Supersymmetry From Universal Extra Dimensions or Little Higgs Models With Dark Matter Experiments}
\author{Dan Hooper and Gabrijela Zaharijas} 
\address{Fermilab, Theoretical Astrophysics Group, Batavia, IL 60510} 
\date{\today}

\bigskip

\begin{abstract} 

There are compelling reasons to think that new physics will appear at or below the TeV-scale.  It is not known what form this new physics will take, however. Although The Large Hadron collider is very likely to discover new particles associated with the TeV-scale, it may be difficult for it to determine the nature of those particles, whether superpartners, Kaluza-Klein modes or other states. In this article, we consider how direct and indirect dark matter detection experiments may provide information complementary to hadron colliders, which can be used to discriminate between supersymmetry, models with universal extra dimensions, and Little Higgs theories. We find that, in many scenarios, dark matter experiments can be effectively used to distinguish between these possibilities.

\end{abstract}

\maketitle
\newpage

\section{Introduction}

As we eagerly await the coming operation of the Large Hadron Collider (LHC), many in the collider phenomenology community are actively scrutinizing what we will be able to learn from such a machine. The TeV energy scale appears extremely likely to contain yet undiscovered physics. The fact that the Higgs mass is quadratically sensitive to new physics suggests that, in order to satisfy constraints from electroweak precision measurements, some kind of new physics must provide a cutoff to these contributions near the TeV scale. The most studied of these possibilities is that of low energy supersymmetry, in which the quadratic divergences from the particles of the Standard Model are canceled by their superpartners, with masses below or not far above 1 TeV. 

Alternatively, so-called Little Higgs theories have been proposed to stabilize the Higgs mass though collective symmetry breaking \cite{little}. In these models, the only contributions to the Higgs mass occur through diagrams involving multiple couplings in the theory, and thus are limited to the 2-loop level and higher. Much like as in supersymmetry, Little Higgs models contain new particles near the TeV scale. Although the precise particle content of Little Higgs theories varies from model to model, they generically contain new gauge bosons, scalars and fermions, each of which have masses near or below 1 TeV \cite{lh}.

Another interesting class of new phenomena at the TeV scale is the appearance of Kaluza-Klein modes, in particular within the context of models with Universal Extra Dimensions (UED) \cite{antoniadis}. In such models, all of the fields of the Standard Model are free to propagate in the bulk of extra dimensional space, which is compactified to a radius of $R \sim$ TeV$^{-1}$. Particles traveling through the higher dimensionality of space appear to four dimensional observers as heavy copies of Standard Model particles -- Kaluza-Klein modes.

If any of these frameworks are manifest in nature, the LHC will be very likely to discover a number of new particles, whether superpartners, Little Higgs partners, or Kaluza-Klein states. Despite the ability of the LHC to detect such particles, however, it is not at all clear that it will be able to determine enough about their detailed properties to distinguish between these various theories. In particular, a critical confirmation of supersymmetry would be to determine that the newly discovered particles are of differing spin from their Standard Model counterparts. If the spins of newly discovered particles are not determined at the LHC, it will be difficult to distinguish whether they are in fact superparticles, or other entities, such as Kaluza-Klein states, for example.

The problem of distinguishing superpartners from Kaluza-Klein states at the LHC has been given quite some attention recently, in particular within the context of UED~\cite{datta,uedsusylhc}. Determining spins using the lepton charge asymmetry is a promising technique~\cite{barr}, but appears to fail in the case of a quasi-degenerate spectrum of the sort expected in UED models \cite{datta,smillie}. Level 2 Kaluza-Klein modes could potentially be discovered at the LHC if they are lighter than about 2 TeV. Such a detection would bolster the case for UED, although a second Kaluza-Klein photon or $Z$-boson could also be confused with a $Z^{\prime}$ \cite{datta}. Similar challenges would be expected for distinguishing supersymmetry from Little Higgs scenarios. If superpartners, Little Higgs partners or Kaluza-Klein modes are light enough to be studied at a future linear collider, the problem of distinguishing these scenarios will become much easier~\cite{linear}. Until that time, however, understanding the findings of the LHC will likely be quite challenging.

Given the challenges involved in distinguishing between the various varieties of new physics that may be discovered at the LHC, it is important to consider other experimental discriminators that will likely be available. In particular, supersymmetry, UED and Little Higgs scenarios can each provide a stable, weakly-interacting particle, suitable as a dark matter candidate. A wide range of dark matter experiments have been developed to search for such particles \cite{dmreview}. These efforts include direct dark matter experiments which attempt to observe the elastic scatterings of WIMPs with nuclei in a detector, and indirect detection efforts which search for the products of WIMP annihilations, including neutrinos, gamma-rays and anti-matter. 

The signatures in direct and indirect dark matter experiments depend on the particle nature of the WIMP being studied. In this paper, we focus on three specific particle dark matter candidates:

\begin{itemize}

\item For the case of supersymmetry, we consider the lightest neutralino, $\chi^0_1$, within the context of the Minimal Supersymmetric Standard Model (MSSM). We assume that R-parity is conserved, and do not introduce any CP-violating phases.

\item For the case of UED, we consider the 5-dimensional case, and study the first KK excitation of the hypercharge gauge boson, $B^{(1)}$~\cite{ST,feng}. Although we do not wed ourselves to a specific KK spectrum, we assume that radiative corrections do not radically alter the quasi-degenerate spectrum which is generally expected \cite{cms}. We also assume that KK-parity is conserved.

\item In Little Higgs theories, we consider the T-parity conserving model of Cheng and Low~\cite{tparity}. For our dark matter candidate, we consider the lightest new heavy neutral gauge boson, $A_H$.

\end{itemize} 

While each of these scenarios can yield a particle with the features required of a dark matter candidate (the mass and annihilation cross section needed to generate the observed relic abundance, for example), their footprints in dark matter experiments may be sufficiently distinctive to provide substantial discriminating power. In this article, we explore the ability of direct and indirect dark matter experiments to distinguish a neutralino from a stable Little Higgs gauge boson or a Kaluza-Klein dark matter state. We find that in many cases, dark matter observations can effectively differentiate between supersymmetry, Little Higgs and UED models.

\section{Direct Detection of Neutralino, Kaluza-Klein and Little Higgs Dark Matter}
\label{directsec}

Direct dark matter detection experiments, such as CDMS \cite{cdms}, ZEPLIN \cite{zeplin}, EDELWEISS \cite{edelweiss}, CRESST \cite{cresst}, XENON \cite{xenon} and WARP \cite{warp}, are designed to detect dark matter particles through their elastic scattering with nuclei. Although WIMPs scatter with nuclei through scalar (spin-independent) and axial-vector (spin-dependent) interactions, current direct detection experiments are far more sensitive to scalar interactions.

The spin-independent WIMP-nuclei elastic scattering cross section is given by:
\begin{equation}
\label{sig}
\sigma_{\rm{SI}} \approx \frac{4 m^2_X m^2_{T}}{\pi (m_X+m_T)^2} [Z f_p + (A-Z) f_n]^2,
\end{equation}
where $m_T$ is the target nuclei's mass, and $Z$ and $A$ are the atomic number and atomic mass of the nucleus.  $f_p$ and $f_n$ are the WIMP's couplings to protons and neutrons, given by:
\begin{equation}
f_{p,n}=\sum_{q=u,d,s} f^{(p,n)}_{T_q} a_q \frac{m_{p,n}}{m_q} + \frac{2}{27} f^{(p,n)}_{TG} \sum_{q=c,b,t} a_q  \frac{m_{p,n}}{m_q},
\label{feqn}
\end{equation}
where $a_q$ are the WIMP-quark couplings and $f^{(p)}_{T_u} \approx 0.020\pm0.004$,  
$f^{(p)}_{T_d} \approx 0.026\pm0.005$,  $f^{(p)}_{T_s} \approx 0.118\pm0.062$,  
$f^{(n)}_{T_u} \approx 0.014\pm0.003$,  $f^{(n)}_{T_d} \approx 0.036\pm0.008$ and 
$f^{(n)}_{T_s} \approx 0.118\pm0.062$ \cite{nuc}. The first term in Eq.~\ref{feqn} 
corresponds to interactions with the quarks in the target nuclei, whereas the second term corresponds to interactions with the gluons in the target through a quark/squark 
loop diagram. $f^{(p)}_{TG}$ is given by $1 -f^{(p)}_{T_u}-f^{(p)}_{T_d}-f^{(p)}_{T_s} 
\approx 0.84$, and analogously, $f^{(n)}_{TG} \approx 0.83$.

We will now turn to specific particle dark matter candidates, beginning with neutralinos. The neutralino-quark coupling, resulting from both Higgs and squark exchange diagrams, is given by~\cite{scatteraq}:
\begin{eqnarray}
\label{aq}
a^{\rm{SUSY}}_q & = & - \frac{1}{2(m^{2}_{1i} - m^{2}_{\chi})} Re \left[
\left( X_{i} \right) \left( Y_{i} \right)^{\ast} \right] 
- \frac{1}{2(m^{2}_{2i} - m^{2}_{\chi})} Re \left[ 
\left( W_{i} \right) \left( V_{i} \right)^{\ast} \right] \nonumber \\
& & \mbox{} - \frac{g_2 m_{q}}{4 m_{W} B} \left[ Re \left( 
\delta_{1} [g_2 N_{12} - g_1 N_{11}] \right) D C \left( - \frac{1}{m^{2}_{H}} + 
\frac{1}{m^{2}_{h}} \right) \right. \nonumber \\
& & \mbox{} +  Re \left. \left( \delta_{2} [g_2 N_{12} - g_1 N_{11}] \right) \left( 
\frac{D^{2}}{m^{2}_{h}}+ \frac{C^{2}}{m^{2}_{H}} 
\right) \right],
\end{eqnarray}
where
\begin{eqnarray}
X_{i}& \equiv& \eta^{\ast}_{11} 
        \frac{g_2 m_{q}N_{1, 5-i}^{\ast}}{2 m_{W} B} - 
        \eta_{12}^{\ast} e_{i} g_1 N_{11}^{\ast}, \nonumber \\
Y_{i}& \equiv& \eta^{\ast}_{11} \left( \frac{y_{i}}{2} g_1 N_{11} + 
        g_2 T_{3i} N_{12} \right) + \eta^{\ast}_{12} 
        \frac{g_2 m_{q} N_{1, 5-i}}{2 m_{W} B}, \nonumber \\
W_{i}& \equiv& \eta_{21}^{\ast}
        \frac{g_2 m_{q}N_{1, 5-i}^{\ast}}{2 m_{W} B} -
        \eta_{22}^{\ast} e_{i} g_1 N_{11}^{\ast}, \nonumber \\
V_{i}& \equiv& \eta_{22}^{\ast} \frac{g_2 m_{q} N_{1, 5-i}}{2 m_{W} B}
        + \eta_{21}^{\ast}\left( \frac{y_{i}}{2} g_1 N_{11},
        + g_2 T_{3i} N_{12} \right)
\label{xywz}
\end{eqnarray}
where throughout $i=1$ for up-type quarks and $i=2$ for down type quarks. $m_{1i}, m_{2i}$ denote elements of the appropriate 2 x 2 squark mass matrix and $\eta$ is the matrix which diagonalizes that matrix. $y_i$, $T_{3i}$ and $e_i$ denote hypercharge, isospin and electric charge of the quarks. For scattering off of up-type quarks:
\begin{eqnarray}
\delta_{1} = N_{13},\,\,\,\, \delta_{2} = N_{14}, \,\,\,\, B = \sin{\beta},\,\,\,\, C = \sin{\alpha}, \,\,\,\, D = \cos{\alpha},
\end{eqnarray}
whereas for down-type quarks:
\begin{eqnarray}
\delta_{1} = N_{14},\,\,\,\, \delta_{2} = -N_{13}, \,\,\,\, B = \cos{\beta},\,\,\,\, C = \cos{\alpha}, \,\,\,\, D = -\sin{\alpha}.
\end{eqnarray}
Here, $\alpha$ is the Higgs mixing angle.

For the case of Kaluza-Klein dark matter, $B^{(1)}$, the coupling is given by \cite{tait}:
\begin{eqnarray}
a^{\rm{UED}}_q  =   \frac{m_q \, g^2_1 \, (Y^2_{q_R}+Y^2_{q_L}) \,  ( m^{2}_{B^{(1)}} + m^{2}_{q^{(1)}})}{4 m_{B^{(1)}}(m^{2}_{B^{(1)}} - m^{2}_{q^{(1)}})^2}  + \frac{m_q \, g^2_1}{8 m_{B^{(1)}}\, m^2_h}, 
\end{eqnarray}
corresponding to contributions from KK quark exchange, and Higgs exchange diagrams, respectively. The first term in this expression is only included for the light quarks ($q=u, d, s$), while the second term contributes for both light and heavy species.

Finally, in the Little Higgs case, the coupling is given by \cite{lhdark}:
\begin{eqnarray}
a^{\rm{LT}}_q  =    \frac{m_q \, g^2_1 \, \tilde{Y}^2 \,  ( m^{2}_{A_H} + m^{2}_{\tilde{Q}})}{4 m_{A_H}(m^{2}_{A_H} - m^{2}_{\tilde{Q}})^2}  +  \frac{m_q \, g^2_1}{8 m_{A_H} \, m^2_h}. 
\end{eqnarray}
Although this is similar to the UED case, the Standard Model hypercharge values are replaced with the hypercharge of the left-handed copy of the Standard Model quarks, $\tilde{Y}=1/10$. Additionally, the mass splitting between the $A_H$ and $\tilde{Q}$ are generally larger than between the $B^{(1)}$ and $q^{(1)}$ in UED, leading to further suppression. For these reasons, scattering in UED is generally dominated by KK quark exchange, while in Little Higgs models, the Higgs exchange contribution is typically larger.

Although the complete expression for the neutralino's elastic scattering cross section is rather complicated, there are a few simple limiting cases of particular interest. First, the largest cross sections are found in the case of large $\tan \beta$ and light $m_H$, and are the result of $H$ exchange, coupled to down-type quarks. For a bino or higgsino-like neutralino and $\cos \alpha \sim 1$, this contribution is approximately given by \cite{carena}:
\begin{eqnarray}
\sigma_{\chi N} &\sim & \frac{g^2_1 g^2_2 |N_{11}|^2 |N_{13}|^2 \,m^4_N \tan^2 \beta}{4\pi m^2_W \, m^4_H} \bigg(f_{T_s}+\frac{2}{27}f_{TG}\bigg)^2 \\ \nonumber &\sim & 4 \times 10^{-7} \rm{pb}   \bigg(\frac{ |N_{11}|^2}{0.9}\bigg)  \bigg(\frac{ |N_{13}|^2}{0.1}\bigg)  \bigg(\frac{\tan \beta}{50}\bigg)^2 \,  \bigg(\frac{300 \, \rm{GeV}}{m_H}\bigg)^4.
\label{susycase1}
\end{eqnarray}
If $H$ is heavy and/or $\tan \beta$ is small, $h$ exchange can dominate. The contribution from $h$ exchange is approximately given by:
\begin{eqnarray}
\sigma_{\chi N} & \sim & \frac{g^2_1 g^2_2 |N_{11}|^2 |N_{14}|^2 \, m^4_N}{4\pi m^2_W \, m^4_h} \bigg(f_{T_u}+\frac{4}{27}f_{TG}\bigg)^2, \\ \nonumber & \sim &  4 \times 10^{-9} \rm{pb}   \bigg(\frac{ |N_{11}|^2}{0.9}\bigg)  \bigg(\frac{ |N_{14}|^2}{0.1}\bigg)  \bigg(\frac{120 \, \rm{GeV}}{m_h}\bigg)^4.
\label{susycase2}
\end{eqnarray}
The contributions from squark exchange diagrams can also be significant, especially if the squarks are only slightly more heavy than the lightest neutralino. The contribution from squark exchange diagrams is approximately given by:%
\begin{eqnarray}
\sigma_{\chi N} &\sim & \frac{3 g^2_1 g^2_2 |N_{11}|^2 |N_{13}|^2 \,m^4_N \tan^2 \beta}{4\pi m^2_W \, (m^2_{\tilde{q}}-m^2_{\chi})^2} \bigg(f_{T_s}+\frac{2}{27}f_{TG}\bigg)^2 \\ \nonumber &\sim & 9 \times 10^{-8} \rm{pb}   \bigg(\frac{ |N_{11}|^2}{0.9}\bigg)  \bigg(\frac{ |N_{13}|^2}{0.1}\bigg)  \bigg(\frac{\tan \beta}{50}\bigg)^2 \,  \bigg(\frac{1 \, \rm{TeV}}{m_{\chi}}\bigg)^4 \, \bigg(\frac{0.1}{\Delta}\bigg)^2,
\label{susycase3}
\end{eqnarray}
where $\Delta \equiv (m_{\tilde{q}}-m_{\chi})/m_{\chi}$.\footnote{Although the masses of the squarks and the lightest neutralino are not typically quasi-degenerate in supersymmetric models, in the models most difficult to distinguish from UED this can be the case.}

In contrast, the cross section for Kaluza-Klein dark matter elastic scattering is approximately given by:
\begin{eqnarray}
\sigma_{B^{(1)} N}   & \approx & \frac{g^4_1 \, m^4_N}{16 \pi m^2_{B^{(1)}}} \bigg[ \frac{1}{m^2_h} \bigg(f_{T_s}+\frac{6}{27}f_{TG}\bigg) + \frac{1}{\Delta^2 m^2_{B^{(1)}}} \bigg(\frac{289}{81} f_{T_u}+ \frac{25}{81}f_{T_s}\bigg)\bigg]^2, \\ \nonumber & \sim &  1.2 \times 10^{-10} \, {\rm pb}\,\bigg(\frac{1\,\rm{TeV}}{m_{B^{(1)}}}\bigg)^2 \, \bigg[\bigg(\frac{100\, \rm{GeV}}{m_h}\bigg)^2 + 0.09 \, \bigg(\frac{1\,\rm{TeV}}{m_{B^{(1)}}}\bigg)^2 \bigg(\frac{0.1}{\Delta}\bigg)^2\bigg]^2,
\label{sigmaueddirect}
\end{eqnarray}
where $\Delta \equiv (m_{q^{(1)}}-m_{B^{(1)}})/m_{B^{(1)}}$. Finally, for Little Higgs dark matter:
\begin{eqnarray}
\sigma_{A_H N}   & \approx & \frac{g^4_1 \, m^4_N}{16 \pi m^2_{A_H}} \bigg[ \frac{1}{m^2_h} \bigg(f_{T_s}+\frac{6}{27}f_{TG}\bigg) + \frac{1}{\Delta^2 m^2_{B^{(1)}}} \bigg(\frac{f_{T_s}}{100}\bigg)\bigg]^2, \\ \nonumber & \sim &  1.2 \times 10^{-10} \, {\rm pb}\,\bigg(\frac{1\,\rm{TeV}}{m_{A_H}}\bigg)^2 \, \bigg[\bigg(\frac{100\, \rm{GeV}}{m_h}\bigg)^2 + 0.0011 \, \bigg(\frac{1\,\rm{TeV}}{m_{A_H}}\bigg)^2 \bigg(\frac{0.1}{\Delta}\bigg)^2\bigg]^2.
\end{eqnarray}
In the Little Higgs case, $\Delta \equiv (m_{\tilde{Q}}-m_{A_H})/m_{A_H}$.

Comparing the expressions for the elastic scattering cross sections in these three scenarios, we notice a few important features. Firstly, the contribution from Higgs exchange in the UED and Little Higgs cases can be approximately related to the contribution from light Higgs ($h$) exchange in supersymmetry by:
\begin{equation}
\frac{\sigma_{\chi N}}{\sigma_{B^{(1)}/A_H \, N}} \sim \frac{16}{9}\frac{g^2_2}{g^2_1} \frac{m^2_{B^{(1)}/A_H}}{m^2_W} |N_{11}|^2 |N_{14}|^2 \sim 80 \, \bigg(\frac{m_{B^{(1)}/A_H}}{1 \, \rm{TeV}}\bigg)^2 \, \bigg(\frac{ |N_{11}|^2}{0.9}\bigg)  \bigg(\frac{ |N_{14}|^2}{0.1}\bigg).
\label{ratio}
\end{equation}
So, even if heavy Higgs ($H$) exchange is negligible, we expect a larger cross section in supersymmetry than in the case of UED or Little Higgs dark matter unless $m_{B^{(1)}/A_H}$ is light and/or the lightest neutralino is a highly pure bino or higgsino. Of course if $\tan \beta$ has a large or moderate value and $m_H$ is not very heavy, then $H$ exchange will dominate over $h$ exchange for neutralino scattering, leading to an even larger ratio than found in Eq.~\ref{ratio}.

We can imagine that this conclusion could be modified in the case of Kaluza-Klein dark matter if the splittings between the KK quark masses and the LKP mass were very small. If $\Delta = 10\%$ and $m_{B^{(1)}}= 500$ GeV, for example, the contribution to the elastic scattering cross section from KK exchange becomes comparable to that from Higgs exchange, leading to a somewhat smaller ratio of approximately $\sigma_{\chi N}/\sigma_{B^{(1)}N} \sim 10 \, (|N_{11}|^2/0.9)(|N_{14}|^2/0.1)$.

\begin{figure}[t]
\centering\leavevmode
\includegraphics[width=4.2in,angle=-90]{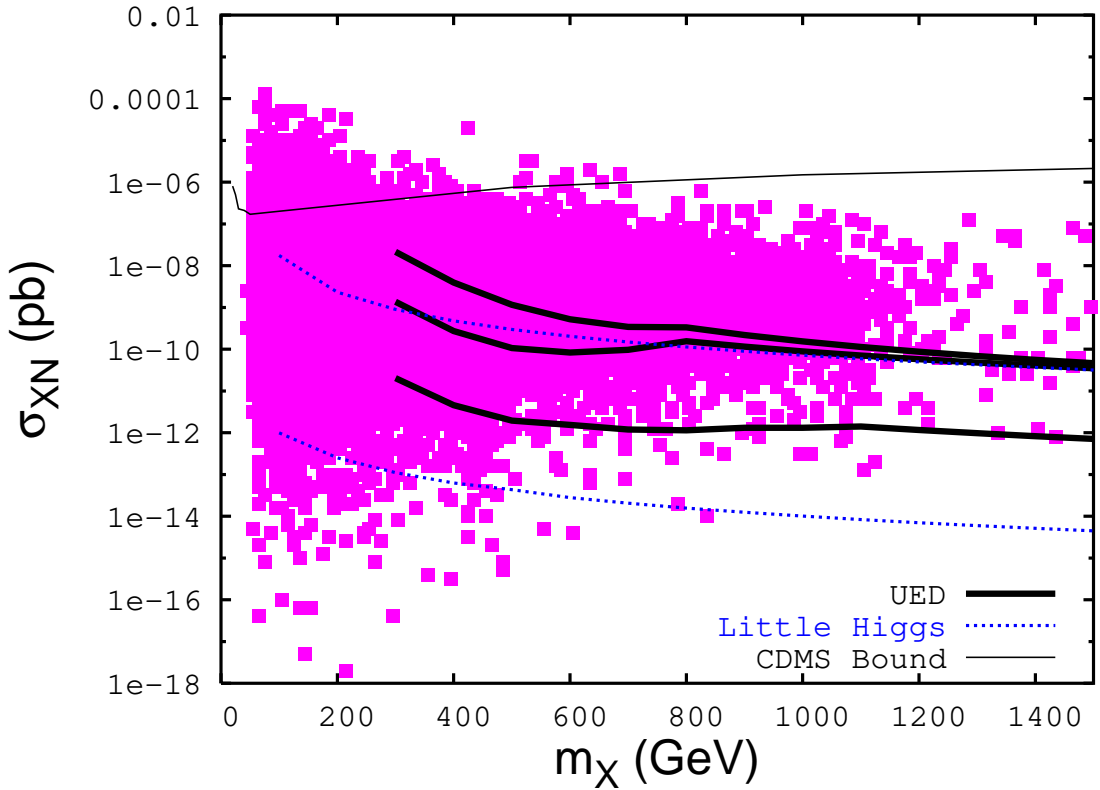}
\caption{A comparison of the WIMP-nucleon, spin-independent elastic scattering cross section in various scenarios. The shaded region represents the range of supersymmetric models found which satisfy all direct collider constraints, the constraints on the $B \rightarrow X_s \gamma $ branching fraction, and do not overproduce the abundance of neutralino dark matter. The thick solid lines denote the UED case for a broad range of KK quark and Higgs masses. From top-to-bottom, the curves denote $m_{q^{(1)}}=1.05 \, m_{B^{(1)}}$ and the minimum Higgs mass, $m_{q^{(1)}}=1.1 \, m_{B^{(1)}}$ and the minimum Higgs mass, and $m_{q^{(1)}}=1.3 \, m_{B^{(1)}}$ and the maximum Higgs mass, respectively (see text for more details). The dotted lines are for the case of T-parity conserving Little Higgs models, with $m_{\tilde{Q}}=1.05\, m_{A_H}$ and $m_h=114$ GeV (top) and with heavy $\tilde{Q}$ and $m_h=1$ TeV (bottom). The thin solid line is the current constraint from the CDMS experiment \cite{cdms}. Note that although we have plotted the cross sections over the full range of WIMP masses, relic abundance considerations limit the LKP mass in UED models to $\gsim 500$ GeV \cite{uedrelic,ST}.}
\label{dir}
\end{figure}

To further study the case of neutralino dark matter, we have performed a scan over the following supersymmetric parameters: $M_2$, $\mu$, sfermion masses, $m_A$, $\tan \beta$ and the trilinear couplings. Each of the masses was allowed to be as large as 4 TeV, and $\tan \beta$ as large as 60.  The range of neutralino masses and spin-independent elastic scattering cross sections with nucleons found in this scan are shown as a shaded region in Fig.~\ref{dir}. This region shows the range of models found which do not violate any direct collider constraints, violate the constraints on the $B \rightarrow X_s \gamma$ branching fraction, and do not overproduce the abundance of neutralino dark matter, as measured by WMAP \cite{wmap}. From this scan, we find that the neutralino-nucleon cross section can vary by more than 10 orders of magnitude. The lower bound to this result is largely a consequence of the maximum values we have allowed $\mu$ and the squark masses to take on in our parameter scan. Models with larger (and more unnatural) values of $\mu$ and/or squark masses would be expected to have even smaller elastic cross sections than shown in the figure.

In Fig.~\ref{dir}, we also show the spin-independent elastic scattering cross sections of WIMPs in UED and Little Higgs models. As expected, the results are somewhat different than in the case of neutralino dark matter. For the case of UED, we have plotted (as thick solid lines) rather extreme cases. The lowest most curve uses KK quark masses 30\% larger than the LKP mass and a Higgs mass as large as is allowed by electroweak precision data (varying from $m_h \approx 900$ GeV for $m_{B^{(1)}}=300$ GeV to $m_h \approx 300$ GeV for $m_{B^{(1)}} \gsim 1$ TeV) \cite{ewpo}. The middle curve was calculated using KK quark masses 10\% larger than the LKP mass and a Higgs mass as low as is allowed by electroweak precision data (varying from $m_h \approx 900$ GeV for $m_{B^{(1)}}=300$ GeV to $m_h = 114$ GeV for $m_{B^{(1)}} \gsim 800$ GeV) \cite{ewpo}. These two curves can be thought of as reasonable upper and lower limits to the spin-independent elastic scattering cross section of the $B^{(1)}$ with nucleons. If we imagine that the KK quarks might be even more degenerate with the LKP, then the cross section could be somewhat larger. The upper thick solid line again uses the minimum possible Higgs mass, but only a 5\% splitting between the KK quark and LKP masses.

The elastic scattering cross section of $A_H$ in T-parity conserving Little Higgs models are shown as dotted lines in Fig.~\ref{dir}. The upper line is for the case of $\tilde{Q}$ 5\% heavier than $A_H$ and a Higgs mass of 114 GeV. The lower line is for the case of a heavy (decoupled) $\tilde{Q}$ and a 1 TeV Higgs mass.

Considering the range of possible cross sections we have calculated in these models, we see that in some cases it may be possible to use the rate in a direct dark matter detection experiment to differentiate supersymmetry from UED or Little Higgs scenarios. In many supersymmetric models, for example, planned direct dark matter detection experiments will measure an elastic scattering cross section above the range predicted in UED or Little Higgs models. In particular, supersymmetric models with large $\tan \beta$ or light $m_H$ generally produce direct detection rates above the range possible in UED or Little Higgs theories. Also, neutralinos with a mixed gaugino-higgsino composition can often generate very high direct detection rates. On the other hand, there is a range of cross sections in which these models will not be distinguishable by direct detection experiments alone. In many of these models, however, combining information from direct and indirect dark matter experiments can enable one to discriminate between supersymmetry, UED and Little Higgs theories.

\section{Neutrinos From Dark Matter Annihilations In The Sun}

WIMPs become captured in the gravitational potential of the Sun at a rate given by \cite{capture}:
\begin{equation} 
C_{\odot} \approx 3.35 \times 10^{18} \, \mathrm{s}^{-1} 
\left( \frac{\rho_{\mathrm{local}}}{0.3\, \mathrm{GeV}/\mathrm{cm}^3} \right) 
\left( \frac{270\, \mathrm{km/s}}{\bar{v}_{\mathrm{local}}} \right)^3  
\left( \frac{\sigma_{\mathrm{H, SD}} + 2.6 \, \sigma_{\mathrm{H, SI}}
+ 0.175 \, \sigma_{\mathrm{He, SI}}}{10^{-6} \, \mathrm{pb}} \right) 
\left( \frac{1 \, \mathrm{TeV}}{m_X} \right)^2 
\label{c-eq}
\end{equation} 
where $\rho_{\mathrm{local}}$ is the local dark matter density and $\bar{v}_{\mathrm{local}}$ is the local RMS velocity of halo dark matter particles. $\sigma_{\mathrm{H,SD}}$, $\sigma_{\mathrm{H, SI}}$ and $\sigma_{\mathrm{He, SI}}$  are the spin-dependent, WIMP-on-proton (hydrogen), spin-independent, WIMP-on-proton and spin-independent, WIMP-on-helium elastic scattering cross sections, respectively. The factors of $2.6$ and $0.175$ include information on the solar 
abundances of elements, dynamical factors and form factor suppression.

The evolution of the number of WIMPs in the Sun, $N$, is given by:
\begin{equation}
\dot{N} = C_{\odot} - A_{\odot} N^2 \; ,
\end{equation}
where $A_{\odot}$ is the 
annihilation cross section multiplied by the relative WIMP velocity per unit volume, given by:
\begin{equation}
A_{\odot} = \frac{\langle \sigma v \rangle}{V_{\mathrm{eff}}} \;.
\end{equation}
Here, $V_{\mathrm{eff}}$ is the effective volume of the Sun's core, which is
determined by matching the core temperature to
a WIMP's gravitational potential energy at the core
radius.  This was found to be \cite{equ}
\begin{equation}
V_{\rm eff} = 1.8 \times 10^{26} \, \mathrm{cm}^3 
\left( \frac{1 \, \mathrm{TeV}}{m_X} \right)^{3/2} \; .
\end{equation}
This leads to a present annihilation rate of WIMPs in the Sun of:
\begin{equation} 
\Gamma = \frac{1}{2} A_{\odot} N^2 = \frac{1}{2} \, C_{\odot} \, 
\tanh^2 \left( \sqrt{C_{\odot} A_{\odot}} \, t_{\odot} \right) \; 
\end{equation}
where $t_{\odot} \simeq 4.5$ billion years is the age of the solar system.
If $\sqrt{C_{\odot} A_{\odot}} t_{\odot} \gg 1$, then the annihilation and capture rates have reached equilibrium, maximizing the resulting neutrino flux. Capture-annihilation equilibrium is reached if the following condition is met:
\begin{equation}
3.35 \times \, \bigg(\frac{\langle \sigma v \rangle}{3 \times 10^{-26}\,\rm{cm}^3/\rm{s}}\bigg)^{1/2} \,\left( \frac{\sigma_{\mathrm{H, SD}} + 2.6 \, \sigma_{\mathrm{H, SI}}
+ 0.175 \, \sigma_{\mathrm{He, SI}}}{10^{-6} \, \mathrm{pb}} \right)^{1/2} \, \bigg(\frac{1\,\rm{TeV}}{m_X}\bigg)^{1/4} \gg 1.
\end{equation}
WIMP annihilation then proceeds to generate neutrinos either directly or through the cascades of their annihilation products. The resulting muon neutrino spectrum at Earth is 
\begin{equation}
\frac{dN_{\nu}}{dE_{\nu}} =\frac{C_{\odot} F_{\rm{Eq}}}{4 \pi D^2_{\rm{ES}}} \bigg(\frac{dN_{\nu}}{dE_{\nu}}\bigg)^{\rm{Inj}}.
\end{equation}
Here, $F_{\rm{Eq}}$ is the non-equilibrium suppression factor ($\approx 1$ for capture-annihilation equilibrium), $D_{\rm{ES}}$ is the Earth-Sun distance, and $(\frac{dN_{\nu}}{dE_{\nu}})^{\rm{Inj}}$ is the neutrino spectrum injected in the Sun per annihilating WIMP (which depends on the dominant annihilation modes).

A small fraction of these neutrinos interact via charged current in the ice or water near a neutrino telescope, generating energetic muon tracks at a rate of:
\begin{equation}
N_{\rm{events}}\simeq \int \int \frac{dN_{\nu_{{\mu}}}}{dE_{\nu_{{\mu}}}} \frac{d\sigma_{\nu}}{dy}(E_{\nu_{\mu}},y)\,R_{\mu}((1-y)E_{\nu})\,A_{\rm{eff}}\, dE_{\nu_{\mu}} dy,
\end{equation}
where $R_{\mu}(E_{\mu})$ is the distance a muon travels before falling below the energy threshold of the detector, called the muon range, and $A_{\rm{eff}}$ is the effective area of the detector. 

For these neutrinos to be identified, they must overcome the background from atmospheric neutrinos. Above 100 GeV, this corresponds to roughly 80 background muons in the Sun's angular window per square kilometer, per year. Over a decade of observation, a 3$\sigma$ detection at an experiment such as IceCube or KM3 would, therefore, require a rate of $\sim 3 \sqrt{80/10} \sim 8$ per square kilometer, per year. To generate such a rate, somewhat large elastic scattering cross sections are needed, {\it ie.} $\sigma_{\mathrm{H, SD}} + 2.6 \, \sigma_{\mathrm{H, SI}} + 0.175 \, \sigma_{\mathrm{He, SI}} \gsim 10^{-6}$ pb. Although this condition does depend on the dominant annihilation modes and the mass of the WIMP, it is generically true that significantly smaller cross sections do not generate observable rate in planned neutrino telescopes.

The constraints on a WIMP's spin-independent elastic scattering cross section from direct detection experiments is currently near the $10^{-7}$--$10^{-6}$ pb level. If CDMS does not make a positive detection in the near future, it will become very unlikely that spin-independent scattering will generate an observable flux of neutrinos from the Sun \cite{halzen}. Spin-dependent scattering, however, is far less constrained. We will focus on these interactions, in particular those spin-dependent couplings that can lead to $\sigma_{\mathrm{H, SD}} \gsim 10^{-6}$ pb.

Neutralinos scatter through axial-vector (spin-dependent) couplings via two classes of diagrams: t-channel $Z$ exchange and s-channel squark exchange. The contribution from squark exchange is unlikely to produce a potentially observable neutrino flux. $Z$-exchange, however, leads to a spin-dependent cross section proportional to the square of the difference of the lightest neutralino's two higgsino fractions:
\begin{eqnarray}
\sigma^{\rm{SUSY}}_{\mathrm{H,SD}} &=& \frac{3 g_2^4 m_p^2}{16 \pi m_W^4} \,
\left(|N_{13}|^2-|N_{14}|^2   \right)^2   \, \left(  T_{3u} \Delta_u^p +  T_{3d} \Delta_d^p + T_{3s} \Delta_s^p \right)^2 \\ \nonumber 
&\approx& 3.5 \times 10^{-6} \, {\rm pb} \bigg(\frac{|N_{13}|^2-|N_{14}|^2}{0.01}\bigg)^2.
\end{eqnarray}
Here, $\Delta_u^p = 0.78 \pm 0.02$, $\Delta_d^p = -0.48 \pm 0.02$, and $\Delta_s^p = -0.15 \pm 0.07$ are the fraction of spin carried by each quark species~\cite{spinfraction}. This can potentially lead to an observable neutrino flux if $(|N_{13}|^2-|N_{14}|^2)^2 \gsim 0.01$ \cite{halzen}.

In the case of UED, spin-dependent scattering naturally dominates over spin-independent, with the cross section generated via KK quark exchange given by:
\begin{eqnarray}
\sigma^{\rm{UED}}_{\mathrm{H,SD}} &=& \frac{g_1^4 m_p^2}{8 \pi m_{B^{(1)}}^2 
(m_{q^{(1)}_R} - m_{B^{(1)}})^2} \,
\left(  (Y^2_{u_R}+Y^2_{u_L}) \Delta_u^p +  (Y^2_{d_R}+Y^2_{d_L}) (\Delta_d^p + \Delta_s^p) \right)^2 \\ \nonumber 
&\approx& 1.8 \times 10^{-6} \, {\rm pb} \bigg(\frac{1 \,{\rm TeV}}{m_{B^{(1)}}}\bigg)^4 \bigg(\frac{0.1}{\Delta}\bigg)^2.
\end{eqnarray}
where again, $\Delta \equiv (m_{q^{(1)}_R} - m_{B^{(1)}})/m_{B^{(1)}}$. Note that for Kaluza-Klein dark matter, the spin-dependent WIMP-proton cross section is much larger than the spin-independent value (see Eq.~\ref{sigmaueddirect}). The flux of neutrinos produced in the case of UED is also enhanced by the relatively large fraction of $B^{(1)}$ annihilations which produce $\tau^+ \tau^-$ (20--25\%) and $\nu \bar{\nu}$ (1--2\%) \cite{Hooper:2002gs}. Neutralinos, in contrast, annihilate largely to heavy quarks and gauge or Higgs bosons. 

In the Little Higgs case, the spin-dependent cross section is again similar to the UED case, but with suppression due to the smaller hypercharge of the quark copies:
\begin{eqnarray}
\sigma^{\rm{LH}}_{\mathrm{H,SD}} &=& \frac{g_1^4 m_p^2 \,  \tilde{Y}^4}{8 \pi m_{A_H}^2 
(m_{\tilde{Q}} - m_{A_H})^2} \,
\left(\Delta_u^p + \Delta_d^p + \Delta_s^p \right)^2 \\ \nonumber 
&\approx& 5.0 \times 10^{-11} \, {\rm pb} \bigg(\frac{1 \,{\rm TeV}}{m_{A_H}}\bigg)^4 \bigg(\frac{0.1}{\Delta}\bigg)^2.
\end{eqnarray}
Here, $\Delta \equiv (m_{\tilde{Q}} - m_{A_H})/m_{A_H}$. Even for a light $A_H$ in the coannihilation region, this cross section is quite small, making rates in planned neutrino telescopes undetectably small.

\begin{figure}[t]
\centering\leavevmode
\includegraphics[width=4.2in,angle=-90]{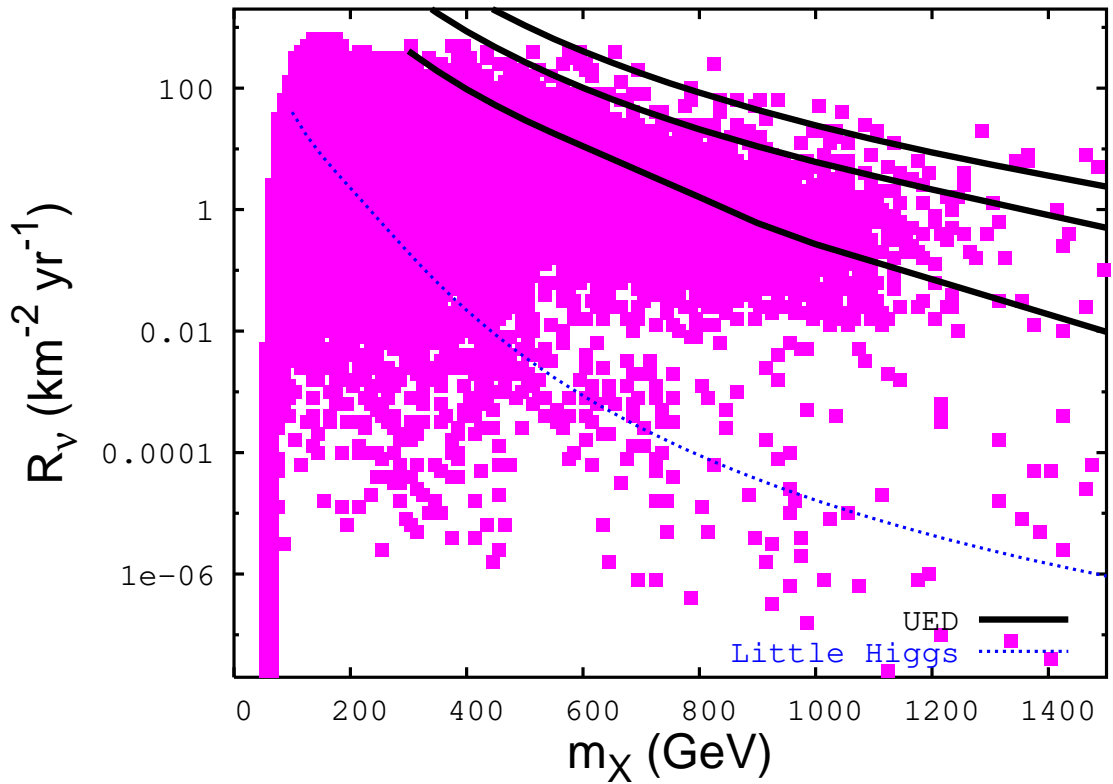}
\caption{A comparison of the rates in neutrino telescopes induced by neutrino produced in dark matter annihilations in the Sun. Again, the shaded region represents the range of supersymmetric models which satisfy all direct collider constraints, the constraints from $B \rightarrow X_s \gamma $, and do not overproduce the abundance of neutralino dark matter. All supersymmetric points shown are also below the current bounds by CDMS \cite{cdms}. The thick solid lines denote the UED case for a broad range of KK quark masses ($m_{q^{(1)}}=1.05 \, m_{B^{(1)}}$, $m_{q^{(1)}}=1.1 \, m_{B^{(1)}}$ and $m_{q^{(1)}}=1.3 \, m_{B^{(1)}}$ from top-to-bottom). The dotted line represents the case of T-parity conserving Little Higgs models, with $m_{\tilde{Q}}=1.05\, m_{A_H}$ and $m_h=114$ GeV. Again note that although we have plotted the cross sections over the full range of WIMP masses, relic abundance considerations limit the LKP mass in UED models to $\gsim 500$ GeV \cite{uedrelic,ST}.}
\label{neu}
\end{figure}

In Fig.~\ref{neu}, we plot the rates in a kilometer-scale neutrino telescope, such as IceCube~\cite{icecube} or KM3~\cite{km3}, from annihilating dark matter in the Sun. From this figure, it is clear that the rate predicted in UED is roughly the largest possible rate from supersymmetry. 

If neutrinos from dark matter annihilations in the Sun were observed, they could certainly be used to rule out the T-parity Little Higgs model we have considered \cite{lhneu} (unless perhaps if $A_H$ is quite light and quasi degenerate with $\tilde{Q}$). On the other hand, the lack of a signal from a WIMP lighter than $\sim$600 GeV (the mass having been determined at the LHC, or in direct detection experiments) could be used to disfavor UED. If a rate consistent with UED were observed, it would be difficult to distinguish from a neutralino with a sizable higgsino component using neutrino telescopes alone.

\section{Positrons From Dark Matter Annihilations In The Local Halo}

WIMP annihilations in the galactic halo generate, among other annihilation products, charged anti-matter particles:  positrons, anti-protons and anti-deuterons. These particles travel through the magnetic fields of our galaxy, diffusing and losing energy, resulting in a diffuse spectrum at Earth. By studying the cosmic anti-matter spectra, satellite-based experiments such as PAMELA \cite{pamela} and AMS-02 \cite{ams02} may be able to identify signatures of dark matter. PAMELA began its three-year mission in June of 2006.

In this section, we will focus on positrons (as opposed to anti-protons or anti-deuterons) for several reasons. Firstly, positrons lose the majority of their energy over typical length scales of a few kiloparsecs or less~\cite{baltzpos}. The cosmic positron spectrum, therefore, samples only the local dark matter distribution and is thus subject to considerably less uncertainty than the other anti-matter species. Secondly, as we will see below, Kaluza-Klein dark matter annihilates via modes that produce a very distinctive positron spectrum \cite{kkpos,feng}. Thirdly, the prospects for studying dark matter with positrons are encouraged by tantalizing, yet ambiguous, hints present in the data from the HEAT experiment \cite{heat}, and to some extent the data from AMS-01 \cite{ams01}.

The spectral shape of the cosmic positron spectrum generated in dark matter annihilation depends on the leading annihilation modes of the WIMP in the low velocity limit. Bino-like neutralinos typically annihilate to heavy fermion pairs: $b\bar{b}$ with a small $\tau^+ \tau^-$ admixture, along with a fraction to $t\bar{t}$ if $m_{\chi} \gsim m_t$. Wino or higgsino-like neutralinos annihilate most efficiently to combinations of Higgs and gauge bosons. 

In UED, Kaluza-Klein dark matter particles annihilate about 60\% of the time to charged lepton pairs, an equal fraction to each of the three families. These directly produced positrons, along with those produced in the decay $\mu^+ \rightarrow e^+ \nu_e \bar{\nu}_{\mu}$ generate a much harder positron spectrum than is expected from neutralinos \cite{kkpos}.

In T-parity conserving Little Higgs scenarios, $A_H$ annihilates largely to $W^+ W^-$ and $ZZ$, with a small ($\lsim 5\%$) admixture to $t\bar{t}$ \cite{lhdark}. Such a particle would be difficult to distinguish from a wino or higgsino-like neutralino from the resulting positron spectrum \cite{lhpos}. 

Once positrons are injected into the local halo through dark matter annihilations, they propagate under the influence of galactic magnetic fields, gradually losing energy through synchrotron emission and inverse Compton scattering with starlight and the cosmic microwave background. We can calculate the spectrum that will be observed at Earth by solving the diffusion-loss equation \cite{diffusion}:
\begin{eqnarray}
\frac{\partial}{\partial t}\frac{dn_{e^{+}}}{dE_{e^{+}}} = \vec{\bigtriangledown} \cdot \bigg[K(E_{e^{+}},\vec{x})  \vec{\bigtriangledown} \frac{dn_{e^{+}}}{dE_{e^{+}}} \bigg] + \frac{\partial}{\partial E_{e^{+}}} \bigg[b(E_{e^{+}},\vec{x})\frac{dn_{e^{+}}}{dE_{e^{+}}}  \bigg] + Q(E_{e^{+}},\vec{x}),
\label{dif}
\end{eqnarray}
where $dn_{e^{+}}/dE_{e^{+}}$ is the number density of positrons per unit energy, $K(E_{e^{+}},\vec{x})$ is the diffusion constant, $b(E_{e^{+}},\vec{x})$ is the rate of energy loss and $Q(E_{e^{+}},\vec{x})$ is the source term, which contains all of the information about the dark matter annihilation modes, cross section and distribution. To solve Eq.~\ref{dif}, a set of boundary conditions must also be adopted. In this application, the boundary condition is described as the distance away from the galactic plane at which the positrons can freely escape, $L$. These diffusion parameters can be constrained by studying the spectra of various species of cosmic ray nuclei, most importantly the boron-to-carbon ratio \cite{btoc}.

\begin{figure}[!]
\centering\leavevmode
\includegraphics[width=3.1in,angle=0]{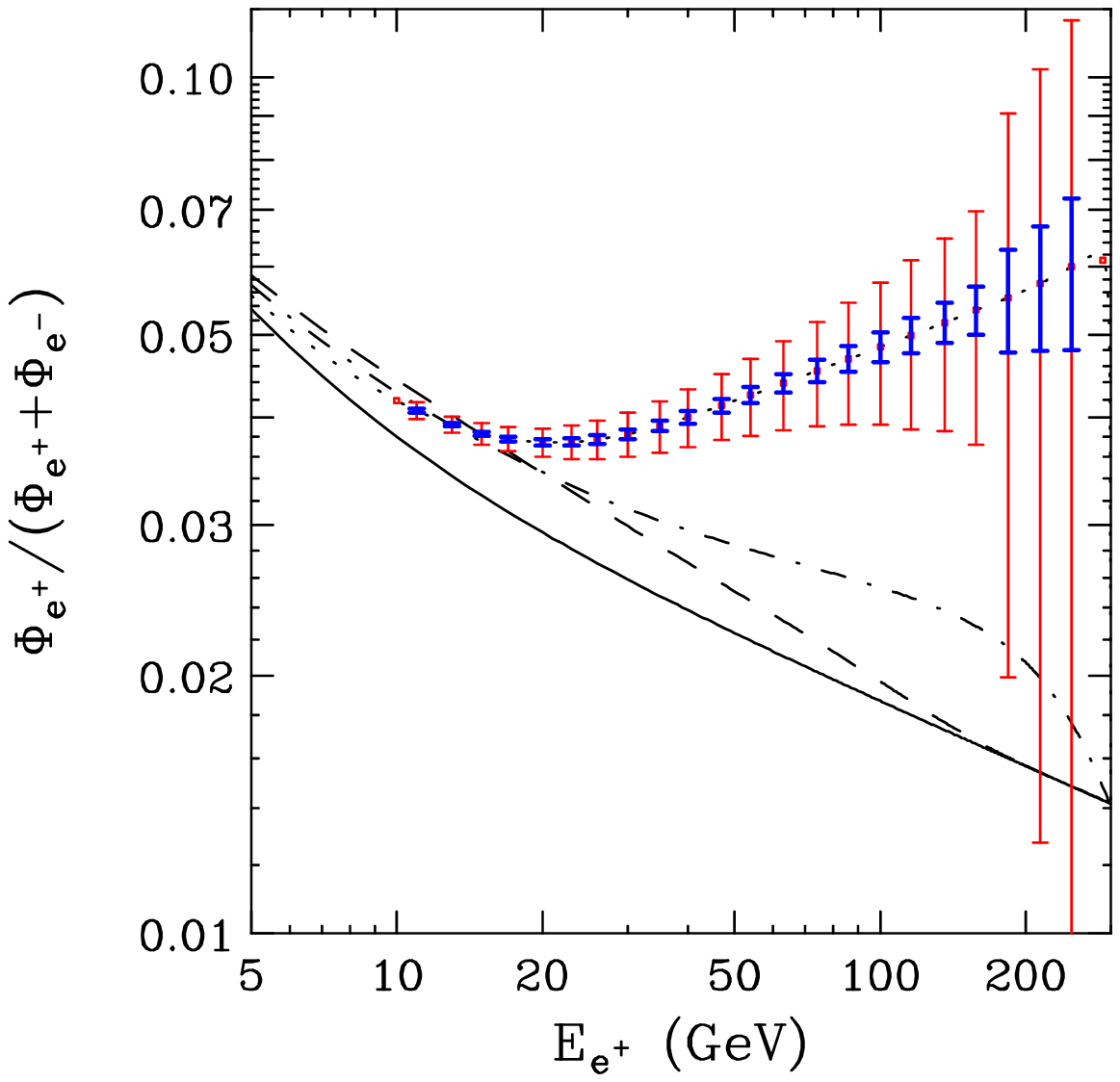} 
\includegraphics[width=3.1in,angle=0]{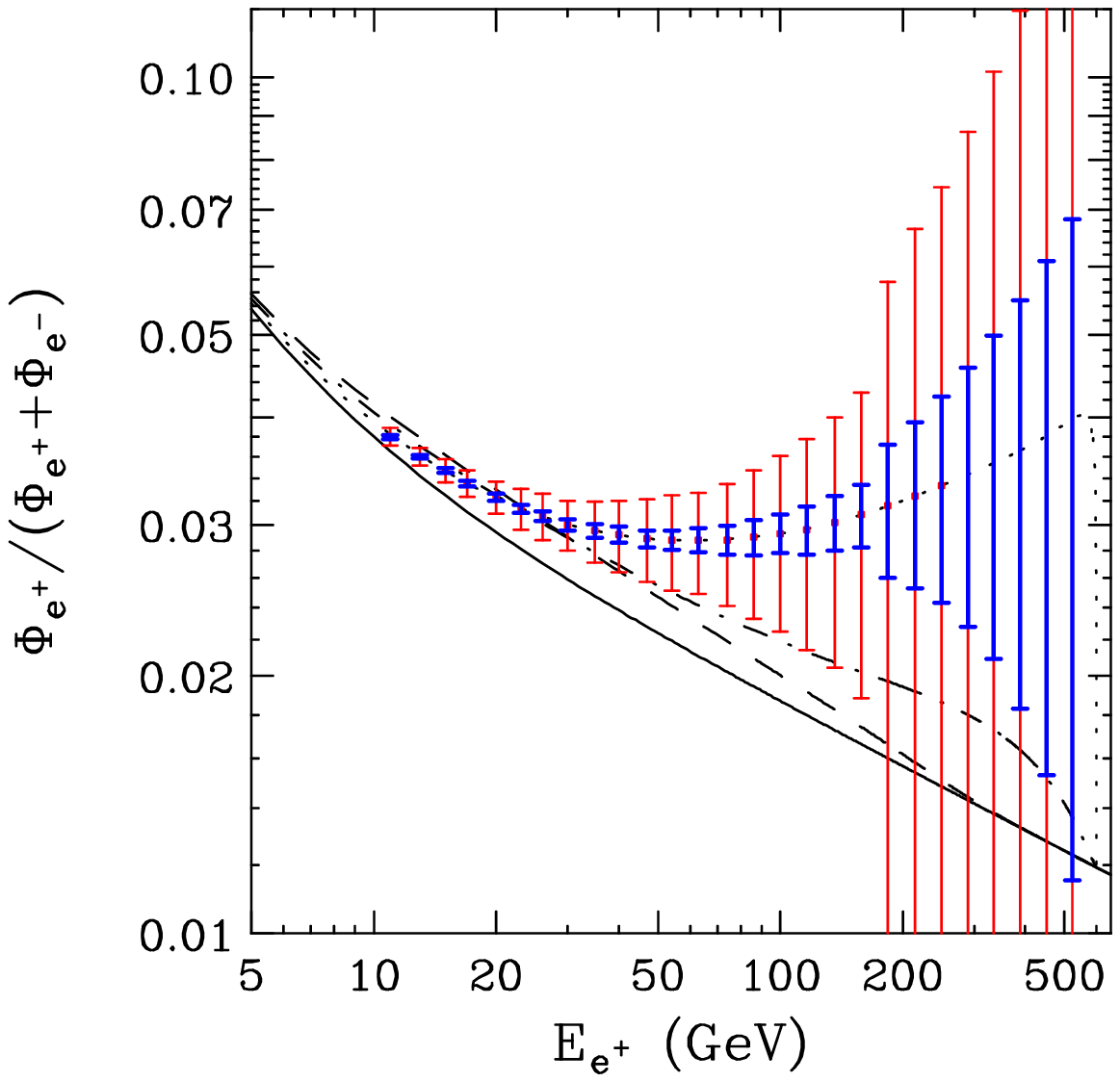} 
\caption{The ratio of positrons to positrons+electrons as a function of energy, after diffusion through the galactic halo, for 300 GeV (left) and 600 GeV (right) WIMPs annihilating to $b\bar{b}$ (dashes) to gauge bosons (dot-dash) and to the modes of Kaluza-Klein (UED) dark matter (dots), which are primarily charged leptons. For each case, we have used the following diffusion parameters: $K(E_{e^+})=3.3 \times 10^{28} E^{0.47}_{e^+}$ cm$^2$/s, $b(E_{e^+})=10^{-16} E^2_{e^+}$ s$^{-1}$ and $L=4$ kpc. To normalize the annihilation rate, in each case we have used a boost factor times annihilation cross section of $30 \times (3 \times 10^{-26})$ cm$^3$/s. The solid line is the astrophysical component from the secondary production of cosmic ray positrons~\cite{secbg}. The error bars shown as those projected for the PAMELA (red, larger) and AMS-02 (blue, smaller) experiments, in the case of Kaluza-Klein (UED) dark matter with $m_{B^{(1)}}=300$ GeV (left), and 600 GeV (right). It is clear that experiments such as these may be able to distinguish UED from supersymmetric or Little Higgs dark matter if the local annihilation rate is sufficiently large.}
\label{posspecEB}
\end{figure}

In Fig.~\ref{posspecEB}, we compare the ratio of positrons to positrons plus electrons in the cosmic ray spectrum, as a function of energy, for different choices of dark matter masses and annihilation modes. We find that the flux of positrons at high energies varies a great deal depending on which type of dark matter particle is considered. To estimate the ability of future cosmic ray measurements to differentiate between the spectra, we have also plotted the error bars projected for the experiments PAMELA and AMS-02 for the case of UED (as found in Ref.~\cite{silkpos}). This clearly demonstrate the ability of these experiments to distinguish Kaluza-Klein dark matter from supersymmetric or Little Higgs models.

Of course, in order for cosmic positron measurements to distinguish between various dark matter candidate, or for that matter to detect dark matter at all, the local dark matter annihilation rate must be sufficiently large. This rate depends on both the annihilation cross section of the WIMP and the degree of local inhomogeneities in the dark matter distribution.

For the simple case of a WIMP annihilating through s-wave contributions ($\sigma v \approx$ constant), without significant coannihilations or resonance effects, a cross section of $\sigma v \approx 3\times 10^{-26}$ cm$^{3}$/s is required to thermally produce the observed dark matter abundance. If terms proportional to $v^2$ are substantial in the annihilation cross section at the temperature of freeze-out, then the low velocity annihilation cross section will have to compensate by being reduced, and the annihilation rate in the local halo will be corresponding suppressed. 

Fluctuations in the local dark matter density can lead to enhancements in the local annihilation rate, known as the ``boost factor''. It is expected that this quantity could be as large as 5 to 10. In order for dark matter annihilations to generate the positron flux observed by HEAT, however, very large boost factors ($\sim 50-100$) are required, or an annihilation cross section well beyond that expected for a thermally produced dark matter relic. If the positron flux observed by HEAT is the result of annihilating dark matter, the spectrum will be precisely measured by PAMELA and AMS-02. If it is not, then the detection of positrons from dark matter annihilations is still possible, but is not guaranteed. The prospects depend strongly on the nature of the WIMP and its distribution. For further discussion on the prospects of PAMELA and AMS-02 to observe positrons from dark matter annihilations, see Ref.~\cite{silkpos}.

In both the UED and Little Higgs cases, the prospects for dark matter detection with positrons is enhanced by the fact that s-wave annihilation terms dominate the freeze-out process ($\sigma v \approx$ constant). This means that the entire annihilation cross section of the WIMP at freeze-out will also be present in the low velocity limit, leading to $\sigma v \sim 3 \times 10^{-26}$ cm$^3$/s in most cases, in the absence of coannihilations. This can also be the case for neutralinos, although this depends on which annihilation channels dominate. Annihilation diagrams through CP-odd Higgs exchange, for example, generate s-wave contributions, while processes such as annihilation through sfermion exchange can lead to $\sigma v \propto v^2$.

\section{Gamma-Rays From Dark Matter Annihilations}

Dark matter annihilations in high density regions could potentially result in an observable flux of gamma\--rays. The galactic center \cite{gc} and satellite dwarf galaxies of the Milky Way \cite{Bergstrom:2005qk,dwarfs} are often considered to be the most promising sites for the detection of such gamma\--rays. Satellite-based (GLAST \cite{glast}) and ground-based (HESS \cite{hess}, MAGIC \cite{magic}, and VERITAS \cite{veritas}) experiments have been designed to search for such a signal.

The differential flux of gamma\--rays produced by dark matter annihilation can be written as, 

\beq
\frac {d\Phi_{\gamma}}{dE_{\gamma}}=\frac {\langle \sigma v \rangle}{2}\frac{1}{4\pi m^2_X}\frac{dN_{\gamma}}{dE_{\gamma}}\int _{los} \rho^2(l) dl, 
\label{flux2}
\eeq
where $m_{X}$ is dark matter mass, $\langle \sigma v \rangle $ is thermally averaged dark matter self-annihilation cross section, $\rho$ is the density of dark matter, and the integral is performed over the line-of-sight. 

The quantity $dN_{\gamma}/dE_{\gamma}$ is the spectrum of gamma-rays produced per annihilation. This spectrum is the result of cascade decays of the dark matter annihilation products, final state radiation~\cite{Bergstrom:2004cy,Birkedal:2005ep}, and line emission ($\gamma \gamma$, $\gamma Z$ or $\gamma h$ final states) \cite{lines}.

We focus first on the processes producing gamma\--rays produced as secondary products. In dark matter annihilation channels to heavy fermions, gauge or Higgs bosons, the resulting gamma\--ray spectrum is quite similar. The exception to this is the somewhat harder spectrum generated through annihilations to $\tau^+ \tau^-$. Neutralinos generally annihilate through $\tau^+ \tau^-$ channel only a few percent of the time or less, however. Regardless of the supersymmetric parameters used, the continuum gamma-ray emission from neutralino dark matter annihilation will be indistinguishable from the spectrum from heavy quarks or gauge/Higgs bosons \cite{andy}. Similarly, Little Higgs dark matter annihilate predominantly into $W$ and $Z$ pairs \cite{lhdark}, and thus cannot be distinguished from neutralinos from the gamma\--ray spectrum.

\begin{figure}
\resizebox{9cm}{!}{\includegraphics{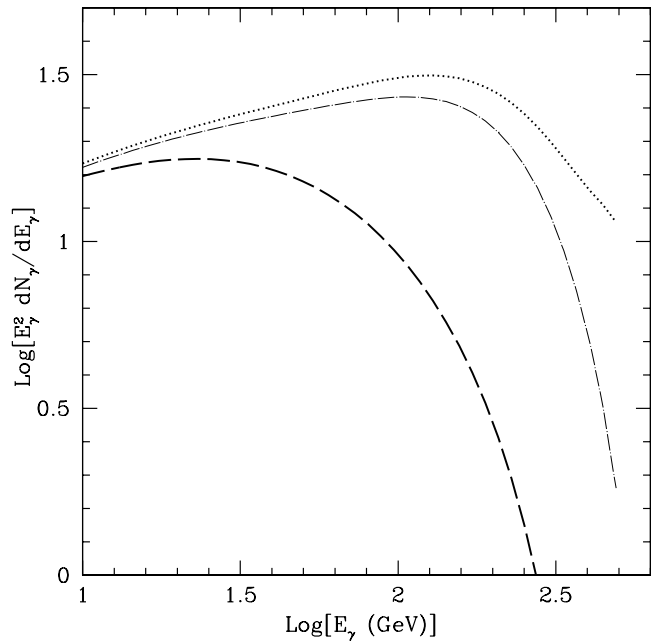}}
\caption{The gamma-ray spectrum produced through the annihilations of Kaluza-Klein dark matter (using $m_{B^{(1)}}=500$ GeV). The dashed line represents the spectrum produced by the fragmentation of quarks and subsequent pion decay. The sum of quark fragmentation and the semi-hadronic $\tau$ decay contributions is plotted as a dot-dashed line. These contributions combined with final state radiation are shown as a dotted line.}
\label{uedspect}
\end{figure}

Kaluza-Klein dark matter, in contrast, annihilates dominantly to charged leptons pairs (20\%--25\% to each generation). This leads to a harder gamma\--ray spectrum than from supersymmetric or Little Higgs dark matter due to the role of tau decays, and final state radiation. These three contributions to the gamma\--ray spectrum from Kaluza-Klein dark matter annihilations are compared in Fig.~\ref{uedspect}. If the gamma-ray spectrum were to be measured with sufficient precision, Kaluza-Klein dark matter could be distinguished from supersymmetric or Little Higgs dark matter.\footnote{Synchotron emission, if observed, could also act as a possible discriminant between KKDM and other dark matter candidates \cite{malcolm}.}

The ability of future experiments to measure the gamma-ray spectrum from dark matter annihilations depends critically on the density of dark matter in regions such as the center of our galaxy, or in dwarf spheroidal galaxies. N-body simulations suggest that dense cusps should be present in such objects, although observations allow for a wide range of possible densities. Baryonic effects, in particular, may lead to the destruction of a dark matter cusp in the center of our galaxy. On the other hand, the effects of adiabatic compression \cite{compression} or the adiabatic accretion of dark matter onto the center supermassive black hole \cite{spike} could lead to extremely high densities in the inner galaxy. 

For the case of the center of our galaxy, we consider two possible dark matter distributions, known as the Navarro-Frenk-White (NFW) \cite{nfw} and Moore {\it et al.} profiles \cite{moore}. The NFW profile features a $\rho(r) \propto r^{-1}$ cusp (where $r$ is the distance to the galactic center) while what we refer to as the Moore {\it et al.} profile features a $\rho(r) \propto r^{-1.4}$ behavior. 

\begin{figure}
\resizebox{8cm}{!}{\includegraphics{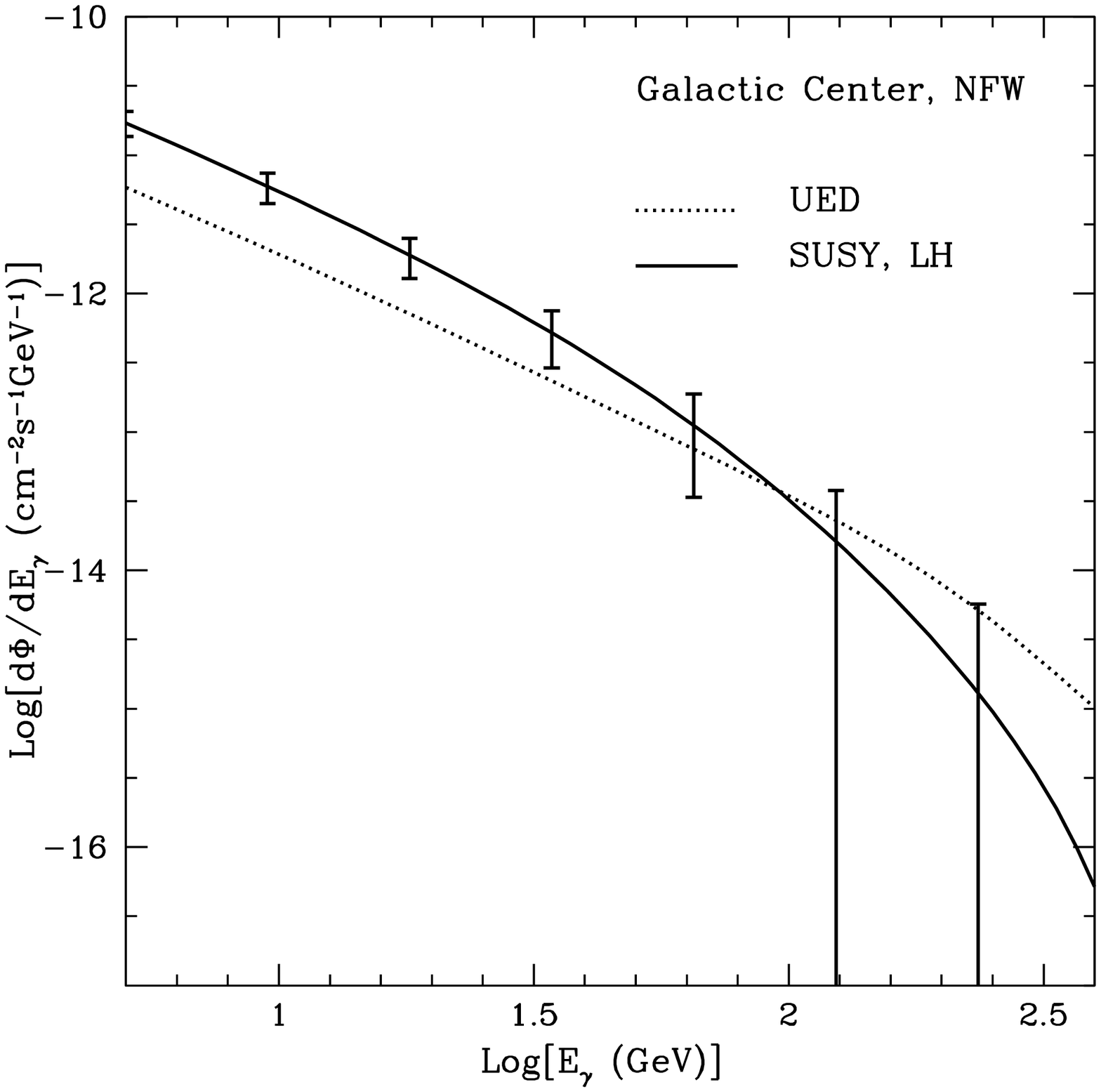}}
\resizebox{8cm}{!}{\includegraphics{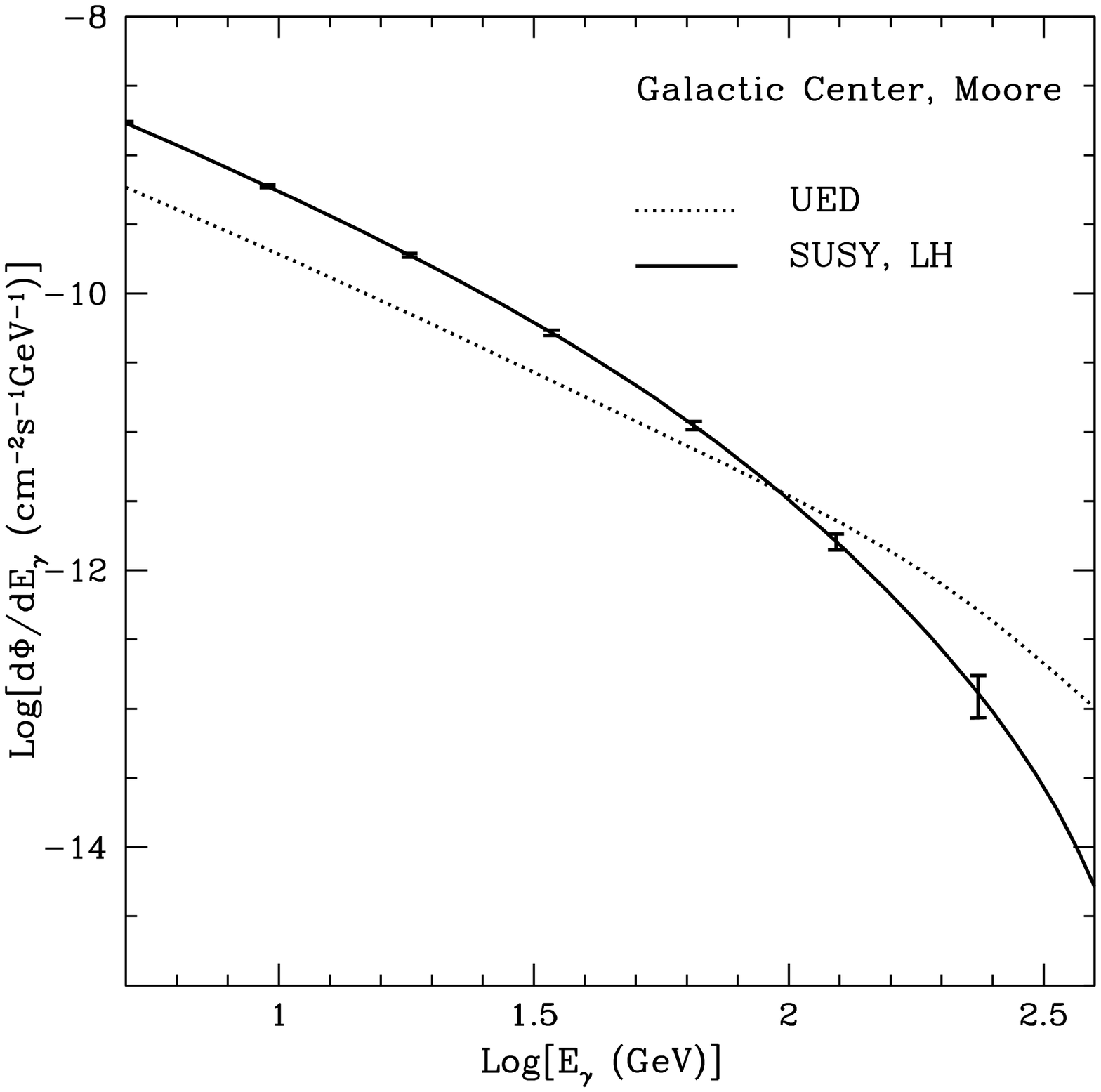}}
\caption{The gamma-ray spectrum from dark matter annihilations in the galactic center region, assuming NFW (left) and Moore {\it et al.} (right) halo profiles. The solid line represents the spectrum for annihilations to gauge bosons (neutralinos or Little Higgs dark matter) while the dotted line is the spectrum from Kaluza-Klein dark matter. The error bars are those projected for the GLAST experiment.  A 500 GeV WIMP with an annihilation cross section of $\sigma v = 3\times 10^{-26}$ cm$^3$/s has been used.}
\label{gc}
\end{figure}

In Fig.~\ref{gc} we show the spectrum of gamma-rays from dark matter annihilations in the galactic center for an NFW (left) and a Moore {\it et al.} (right) profile. In each frame, we have used $\sigma v = 3 \times 10^{26}$ cm$^3$/s and a dark matter mass of 500 GeV. The error bars shown correspond to the projections for the GLAST satellite. 

It would appear from Fig.~\ref{gc} that discriminating between supersymmetric or Little Higgs dark matter and Kaluza-Klein dark matter with GLAST should be possible, even in the case of an NFW halo profile. We have not yet, however, considered the effect of the known astrophysical TeV gamma-ray source located at the galactic center \cite{hess,magic}. Once this is taken into account, an NFW profile will lead to a positive detection of dark matter by GLAST only for a particle with a mass below $\sim 100$ GeV \cite{gabi}. If a more dense profile (such as Moore {\it et al.}) is considered, however, this background could potentially be overcome.

Given the challenges involved with observing dark matter in the galactic center, we will also consider the possibility of detecting gamma-rays from dark matter annihilations in a class of companion galaxies of the Milky Way, called dwarf spheroidals \cite{Bergstrom:2005qk,dwarfs}. Dwarf spheroidal galaxies are highly dark matter dominated, and are not expected to produce significant backgrounds in their inner regions. Draco is one of the closest and most massive dwarf galaxies, and even though it is not expected to be the brightest gamma\--ray source among them, its dark matter density profile is the most tightly constrained by observations, which makes it an interesting candidate for gamma\--ray detection.

The flux of gamma\--rays from dark matter annihilations in Draco was estimated in Ref.~\cite{Bergstrom:2005qk} to be
\beq
\Phi_{\rm{Draco}} \approx 2.4\times 10^{-10} \, \, \rm{to} \, \, 3.5 \times 10^{-13} \, \left( \frac{100 {\rm GeV}}{m_{X}}\right) ^2 \left(\frac {\langle \sigma v \rangle}{3\times 10^{-26} {\rm cm}^3 {\rm s}^{-1}}\right) \left( \frac {N_{\gamma}}{10} \right)~~{\rm cm}^{-2}{\rm s}^{-1},
\label{dracoeq}
\eeq
where $N_{\gamma}$ is number of photons per annihilation emitted in the energy range of a given experiment. The range of fluxes shown here reflects the range of halo profiles consistent with observations.

\begin{figure}
\resizebox{8cm}{!}{\includegraphics{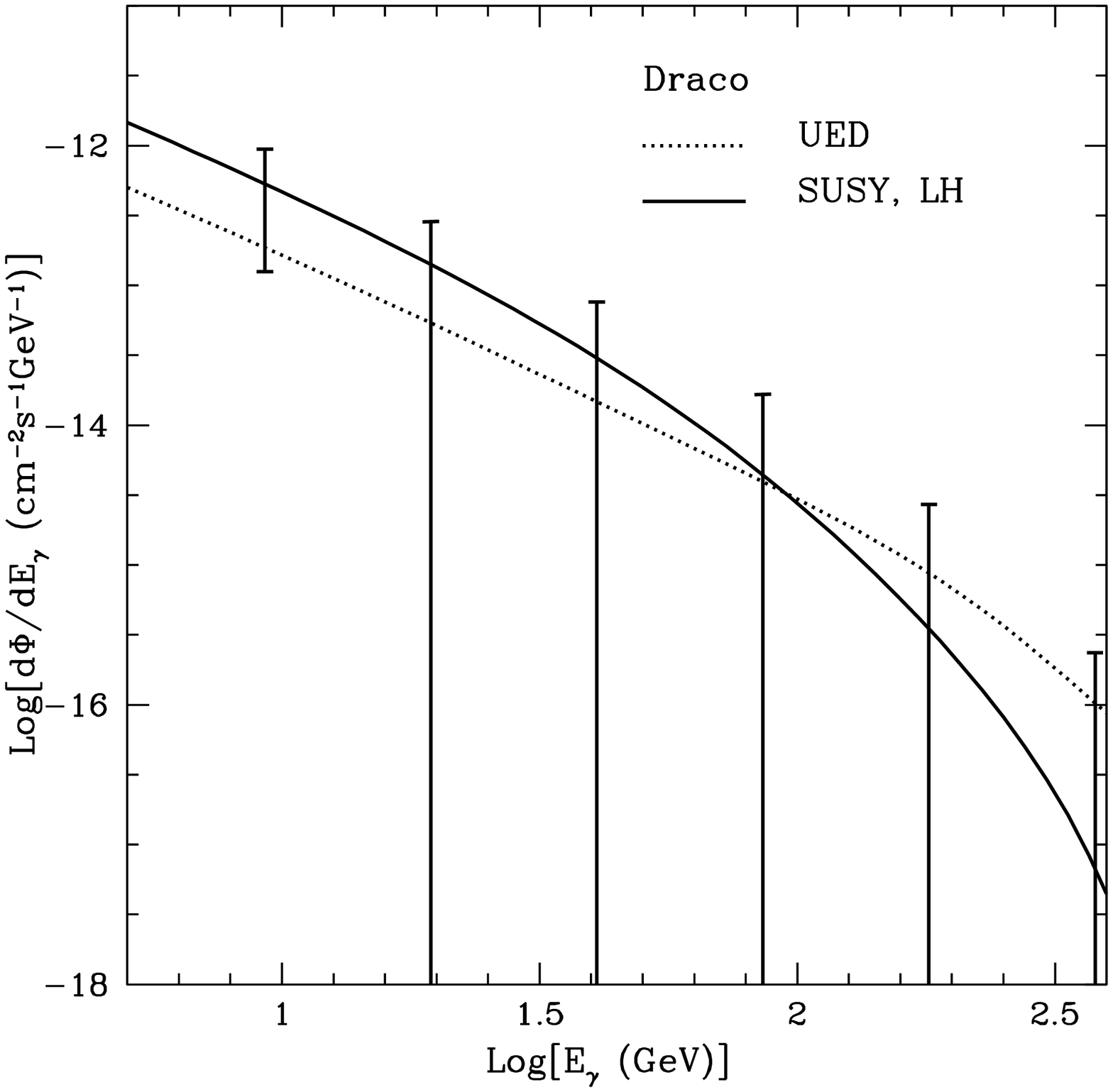}}
\resizebox{8cm}{!}{\includegraphics{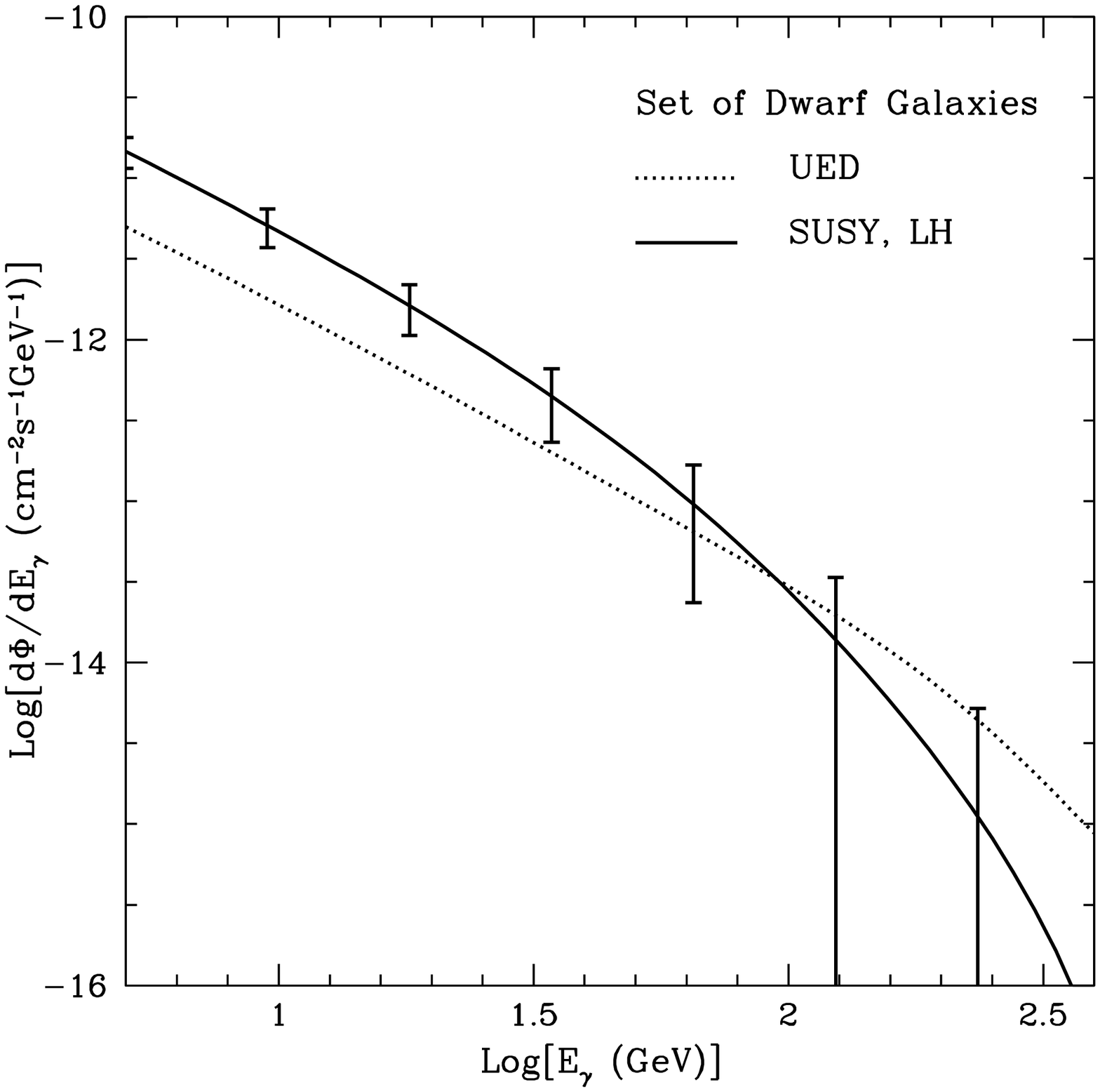}}
\caption{The gamma-ray spectrum from dark matter annihilations in the dwarf spheroidal galaxy Draco (assuming the maximal flux from Eq.~\ref{dracoeq}) is shown in the left frame. In the right frame, we plot a flux ten times larger, speculating that the sum of all dwarf galaxies could generate such a flux. The solid line represents the spectrum for annihilations to gauge bosons (neutralinos or Little Higgs dark matter) while the dotted line is the spectrum from Kaluza-Klein dark matter. The error bars are those projected for the GLAST experiment. A 500 GeV WIMP with an annihilation cross section of $\sigma v = 3\times 10^{-26}$ cm$^3$/s has been used.}
\label{draco}
\end{figure}

In the left frame of Fig.~\ref{draco}, we plot the gamma\--ray spectrum from dark matter annihilation using the maximal flux shown in Eq.~\ref{dracoeq}. A 500 GeV mass and an annihilation cross section of $3 \times 10^{-26}$ cm$^3$/s were used. This flux would lead to a marginal detection by GLAST, but would not provide the level of precision needed to distinguish between supersymmetric, Little Higgs or Kaluza-Klein dark matter. 

Although Draco is the dwarf spheroidal which has its halo profile most tightly constrained by observations, is not necessarily the brightest in dark matter annihilation radiation, and there exist many other dwarfs in the Milky Way which could generate observable fluxes of gamma-rays. In the right frame of Fig.~\ref{draco}, we speculate that the total flux of gamma-rays from all dwarfs is ten times larger than that shown for Draco alone in the left frame. In this case, supersymmetric/Little Higgs dark matter could be distinguished from UED by GLAST.

\section{UED-Like Supersymmetric Models}

In this section, we consider a series of sample supersymmetric models which could be difficult to distinguish from UED at the LHC. It is in this subset of models that the discriminating power of dark matter experiments is the most interesting and potentially useful.

To begin, we consider a number of supersymmetric models which possess, in some respects, the features expected in UED. In table~\ref{uedlike}, we describe a sample of such models. Models 1a-1c resemble a UED model with $R^{-1} \approx 550$ GeV and $\Lambda R \approx 20$ (leading to a KK quark-LKP mass splitting of approximately 20\%). Models 2a-2c resemble UED with $R^{-1} \approx 800$ GeV and a KK quark-LKP mass splitting of approximately 30\%. In the table, we give the values of the input parameters $M_1$, $M_3$, $\mu$, $\tan \beta$, $m_{\tilde{f}}$ and $m_A$. We have set $M_2=2 M_1$ and set the trilinear coupling $A_t$ to minimize the stop mass splitting. Although these models do not contain particle spectra identical to those predicted by UED, there are enough similarities to potentially be indistinguishable at the LHC.
\vspace{0.5cm}
 \begin{table}[!ht]
 \hspace{0.0cm}
 \begin{tabular} {c c c c c c c c c c c c c c c c c c c c c c c} 
\hline \hline
 Model &\vline& $M_1$ & \,\,& $M_3$  & \,\, &  $\mu$ &\,\,&  $\tan \beta$ &\,\,&  $m_{\tilde{f}}$ &  \,\,&  $m_A$ &\vline& $m_{\chi}$ & \,\, &  $\Omega_{\chi^0} h^2$ && $f_{g}$  &\vline& $\sigma_{\rm{SI}}$ & \, \, & $R_{\nu}$ \\
 \hline \hline
1a &\vline& 570. && 700. && 580. && 10. && 650. && 1500. &\vline& 540. && 0.097 && 50\% &\vline& $3.9 \times 10^{-8} \, \rm{pb}$ && 69.\\ 
\hline
1b &\vline& 570. && 700. && 580. && 10. && 650. && 700. &\vline& 540. && 0.082 && 50\% &\vline& $4.9 \times 10^{-8} \, \rm{pb}$ && 67.\\ 
\hline
1c &\vline& 550. && 700. && 700. && 30. && 650. && 700. &\vline& 547. && 0.076 && 95\% &\vline& $1.4 \times 10^{-8} \, \rm{pb}$ && 8.1\\ 
 \hline \hline
2a &\vline& 830. && 1100. && 840. && 10. && 1050. && 1500. &\vline& 802. && 0.106 && 50\% &\vline& $2.8 \times 10^{-8} \, \rm{pb}$ && 15.\\
\hline
2b &\vline& 830. && 1100. && 840. && 10. && 1050. && 700. &\vline& 802. && 0.136 && 50\% &\vline& $3.6 \times 10^{-8} \, \rm{pb}$ && 27.\\ 
\hline
2c &\vline& 830. && 1100. && 840. && 30. && 1050. && 700. &\vline& 804. && 0.117 && 50\% &\vline& $5.6 \times 10^{-8} \, \rm{pb}$ && 37.\\ 
\hline \hline
 \end{tabular}
 \caption{A sample of supersymmetric models which would be difficult to distinguish from UED at the LHC. In each model, $M_2=2 M_1$ and $A_t=A_b=\mu /\tan \beta$. All masses are in GeV. The neutrino rate, $R_{\nu}$, is given in events per year per square kilometer effective area. By combining the rates in direct detection experiments and neutrino telescopes, each of these models can be distinguished from UED.}
\label{uedlike}
 \end{table}

Also shown in the table are the mass, thermal relic abundance and gaugino fraction ($f_g$) of the lightest neutralino. In each case, the composition (through the parameter $\mu$) was chosen to yield a relic abundance similar to the measured abundance of dark matter. Finally, for each UED-like supersymmetric model, we show the spin-independent elastic scattering cross section (as is relevant to direct detection experiments) and the rate in a kilometer-scale neutrino telescope, such as IceCube, per year. For each of the supersymmetric models, we find elastic scattering cross sections of a few times $10^{-8}$ pb. This is the case for two reasons. Firstly, the squark masses and neutralino mass are quasi-degenerate, leading to substantial squark exchange contributions to the elastic scattering cross section (see Eq.~\ref{susycase3}). Secondly, for such heavy neutralinos to generate the observed relic abundance, they must possess a sizable higgsino component. Together, these features lead to elastic scattering cross sections approximately in the range shown in table \ref{uedlike}.

For such a large cross section to be found in UED models, the KK quarks would have to be only a few percent heavier than the LKP (see Fig.~\ref{dir}). Such a small mass splitting in UED would also lead to thousands of events being generated each year in IceCube. As our supersymmetric models each predict far fewer neutrino events, we conclude that the combination of direct detection experiments and neutrino telescope can distinguish between supersymmetry and UED in each of these scenarios.

\begin{figure}[t]
\centering\leavevmode
\includegraphics[width=4.2in,angle=-90]{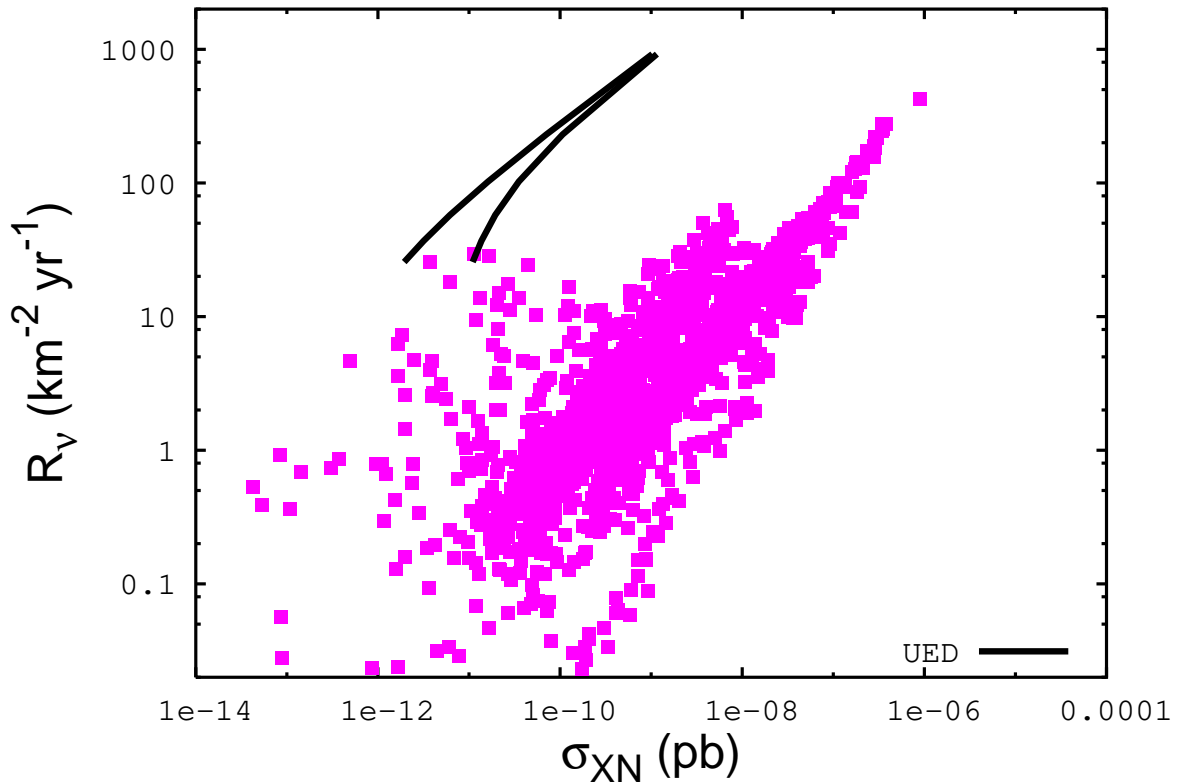}
\caption{A comparison of the neutrino and direct detection rates for neutralinos and Kaluza-Klein dark matter. Again, the shaded region represents the range of supersymmetric models which satisfy all direct collider constraints, the constraints from $B \rightarrow X_s \gamma $, and do not overproduce the abundance of neutralino dark matter. All supersymmetric points shown are also below the current bounds by CDMS \cite{cdms}. Furthermore, in this figure, only models with neutralino masses in the range of 450-550 GeV are shown. The thick solid lines denote the UED case with $m_{B^{(1)}}=500$ GeV, and varied across a broad range of KK quark masses ($m_{q^{(1)}}=1.05 \, m_{B^{(1)}}$ to \, $1.3 \, m_{B^{(1)}}$). The two UED contours represent the maximum and minimum Higgs masses consistent with electroweak precision observables.}
\label{com}
\end{figure}

In Fig.~\ref{com}, we have plotted the results of our earlier supersymmetric parameter scan, comparing the elastic scattering cross section to the rate in neutrino telescopes. Each supersymmetric model shown contains a lightest neutralino with a mass in the range of 450-550 GeV. When the area of this plane occupied by supersymmetric models is compared to the results for a 500 GeV Kaluza-Klein dark matter particle (shown as solid lines), it is clear that the combination of direct detection and neutrino experiments can be a powerful discriminator between supersymmetry and UED.

\section{Summary and Conclusions}

In this paper, we have explored ways in which direct and indirect dark matter experiments could be used to distinguish between supersymmetry, Universal Extra Dimensions (UED) and Little Higgs models. To summarize, we find that:
\begin{itemize}
\item{Spectral features in the cosmic positron spectrum could be used to distinguish UED from supersymmetry or Little Higgs models. This is because Kaluza-Klein dark matter particles, unlike neutralinos or Little Higgs dark matter, annihilate efficiently to light fermions including $e^+ e^-$, $\mu^+ \mu^-$ and $\tau^+ \tau^-$, leading to a very hard positron spectrum compared to dark matter annihilations to heavy quarks or gauge bosons. The usefulness of this channel will depend critically on the unknown dark matter annihilation rates in the Galactic Center and in dwarf spheriodal galaxies.}

\item{In UED, dark matter annihilations to $\tau^+ \tau^-$ lead to a harder gamma-ray spectrum than is found in the case of supersymmetry or Little Higgs. Furthermore, the contribution from final state radiation makes the spectrum from Kaluza-Klein dark matter even more substantial at high energies. The usefulness of this channel will depend critically on the unknown local dark matter annihilation rate.}

\item{The neutralino-nucleon elastic scattering cross sections can, for some parameters, be larger than is found in either UED or Little Higgs models. If such a large cross section is observed, it would strongly disfavor UED and Little Higgs scenarios. On the other hand, if a smaller cross section is measured in future direct detection experiments, it could not easily be used to distinguish between these scenarios (see Fig.~\ref{dir}).}

\item{The flux of neutrinos from dark matter annihilations in the Sun is very large in the case of UED. Dark matter in Little Higgs theories, in contrast, produces a very small neutrino flux from the Sun. The neutrino flux from supersymmetric dark matter can vary over this entire range of possibilities (see Fig.~\ref{neu}).}

\item{By combining information from direct detection and neutrino experiments, it may be possible to achieve further discriminating power. In particular, Kaluza-Klein dark matter results in larger neutrino-to-direct rates than is found for neutralino dark matter.}

\end{itemize}

The ability of dark matter experiments to detect and study dark matter will depend strongly on a number of unknown inputs of both particle and astrophysical natures. In particular, the prospects for studying dark matter with gamma-ray and positron telescopes will depend critically on the distribution of dark matter in our galaxy. 

Although the LHC will almost certainly reveal much to us about the nature of the TeV-scale, it will also almost certainly not tell us everything we would like to know. Dark matter experiments can provide a set of information which is highly complementary to that provided by colliders. In many cases, it may be possible for dark matter experiments to break degenercies between different models which are indistinguishable at the LHC.

\acknowledgements{We would like to acknowledge Graham Kribs and Jay Hubisz for many important discussions related to this work, some of which began at the University of Oregon UltraMini Workshop on Cosmology and Astrophysics.  DH is supported by the Department of Energy and by NASA grant NAG5-10842.}


\begin{thebibliography}{10}



\bibitem{little}
 N.~Arkani-Hamed, A.~G.~Cohen and H.~Georgi,
  Phys.\ Lett.\ B {\bf 513}, 232 (2001)
  [arXiv:hep-ph/0105239].

\bibitem{lh}
  N.~Arkani-Hamed, A.~G.~Cohen, E.~Katz and A.~E.~Nelson,
  JHEP {\bf 0207}, 034 (2002)
  [arXiv:hep-ph/0206021];
  N.~Arkani-Hamed, A.~G.~Cohen, E.~Katz, A.~E.~Nelson, T.~Gregoire and J.~G.~Wacker,
  JHEP {\bf 0208}, 021 (2002)
  [arXiv:hep-ph/0206020];
 I.~Low, W.~Skiba and D.~Smith,
  Phys.\ Rev.\ D {\bf 66}, 072001 (2002)
  [arXiv:hep-ph/0207243];
  D.~E.~Kaplan and M.~Schmaltz,
  JHEP {\bf 0310}, 039 (2003)
  [arXiv:hep-ph/0302049];
  S.~Chang and J.~G.~Wacker,
  Phys.\ Rev.\ D {\bf 69}, 035002 (2004)
  [arXiv:hep-ph/0303001];
  W.~Skiba and J.~Terning,
  Phys.\ Rev.\ D {\bf 68}, 075001 (2003)
  [arXiv:hep-ph/0305302];
  S.~Chang,
  JHEP {\bf 0312}, 057 (2003)
  [arXiv:hep-ph/0306034];
  M.~Schmaltz,
  JHEP {\bf 0408}, 056 (2004)
  [arXiv:hep-ph/0407143].



\bibitem{antoniadis}
I.~Antoniadis,
Phys.\ Lett.\ B {\bf 246}, 377 (1990);
I.~Antoniadis, K.~Benakli and M.~Quiros,
Phys.\ Lett.\ B {\bf 331}, 313 (1994).

\bibitem{datta}
  A.~Datta, K.~Kong and K.~T.~Matchev,
  Phys.\ Rev.\ D {\bf 72}, 096006 (2005)
  [Erratum-ibid.\ D {\bf 72}, 119901 (2005)]
  [arXiv:hep-ph/0509246].


\bibitem{uedsusylhc}
H.~C.~Cheng, K.~T.~Matchev and M.~Schmaltz,
  Phys.\ Rev.\ D {\bf 66}, 056006 (2002)
  [arXiv:hep-ph/0205314];
  A.~Datta, G.~L.~Kane and M.~Toharia,
  arXiv:hep-ph/0510204;
  A.~Alves, O.~Eboli and T.~Plehn,
  Phys.\ Rev.\ D {\bf 74}, 095010 (2006)
  [arXiv:hep-ph/0605067];
  C.~Athanasiou, C.~G.~Lester, J.~M.~Smillie and B.~R.~Webber,
  JHEP {\bf 0608}, 055 (2006)
  [arXiv:hep-ph/0605286].


\bibitem{barr}
  A.~J.~Barr,
  Phys.\ Lett.\ B {\bf 596}, 205 (2004)
  [arXiv:hep-ph/0405052].


\bibitem{smillie}
J.~M.~Smillie and B.~R.~Webber,
  JHEP {\bf 0510}, 069 (2005)
  [arXiv:hep-ph/0507170].




\bibitem{linear}
  M.~Battaglia, A.~Datta, A.~De Roeck, K.~Kong and K.~T.~Matchev,
  JHEP {\bf 0507}, 033 (2005)
  [arXiv:hep-ph/0502041].


\bibitem{dmreview}
  G.~Bertone, D.~Hooper and J.~Silk,
  Phys.\ Rept.\  {\bf 405}, 279 (2005)
  [arXiv:hep-ph/0404175].




\bibitem{tparity}
  H.~C.~Cheng and I.~Low,
  JHEP {\bf 0309}, 051 (2003)
  [arXiv:hep-ph/0308199];
 H.~C.~Cheng and I.~Low,
  JHEP {\bf 0408}, 061 (2004)
  [arXiv:hep-ph/0405243];
  I.~Low,
  JHEP {\bf 0410}, 067 (2004)
  [arXiv:hep-ph/0409025].


\bibitem{ST}
G.~Servant and T.~M.~Tait,
Nucl.\ Phys.\ B {\bf 650}, 391 (2003)
  [arXiv:hep-ph/0206071].


\bibitem{feng}
H.~C.~Cheng, J.~L.~Feng and K.~T.~Matchev,
  Phys.\ Rev.\ Lett.\  {\bf 89}, 211301 (2002)
  [arXiv:hep-ph/0207125].




\bibitem{cms}
 H.~C.~Cheng, K.~T.~Matchev and M.~Schmaltz,
  Phys.\ Rev.\ D {\bf 66}, 036005 (2002)
  [arXiv:hep-ph/0204342].



\bibitem{cdms}
 D.~S.~Akerib {\it et al.}  [CDMS Collaboration],
  Phys.\ Rev.\ Lett.\  {\bf 96}, 011302 (2006)
  [arXiv:astro-ph/0509259];
 D.~S.~Akerib {\it et al.}  [CDMS Collaboration],
  Phys.\ Rev.\ D {\bf 73}, 011102 (2006)
  [arXiv:astro-ph/0509269].

 
\bibitem{zeplin}
  G.~J.~Alner {\it et al.}  [UK Dark Matter Collaboration],
  Astropart.\ Phys.\  {\bf 23}, 444 (2005).

\bibitem{edelweiss}
  V.~Sanglard {\it et al.}  [The EDELWEISS Collaboration],
  Phys.\ Rev.\ D {\bf 71}, 122002 (2005)
  [arXiv:astro-ph/0503265].


\bibitem{cresst}
  G.~Angloher {\it et al.},
  Astropart.\ Phys.\  {\bf 23}, 325 (2005)
  [arXiv:astro-ph/0408006].

\bibitem{xenon}
  T.~Shutt, C.~E.~Dahl, J.~Kwong, A.~Bolozdynya and P.~Brusov,
  arXiv:astro-ph/0608137.


\bibitem{warp}
  R.~Brunetti {\it et al.},
  New Astron.\ Rev.\  {\bf 49}, 265 (2005)
  [arXiv:astro-ph/0405342];
Preliminary limits using liquid Argon were shown at the April 2006, P5 meeting at Fermilab by Elena Aprile (http://www.fnal.gov/directorate/program$_{-}$planning/P5/P5$_{-}$Apr2006/Talks/Aprile.pdf).



\bibitem{nuc}
  A.~Bottino, F.~Donato, N.~Fornengo and S.~Scopel,
  Astropart.\ Phys.\  {\bf 18}, 205 (2002)
  [arXiv:hep-ph/0111229];
 Astropart.\ Phys.\  {\bf 13}, 215 (2000)
  [arXiv:hep-ph/9909228];
  J.~R.~Ellis, K.~A.~Olive, Y.~Santoso and V.~C.~Spanos,
  Phys.\ Rev.\ D {\bf 71}, 095007 (2005)
  [arXiv:hep-ph/0502001].

\bibitem{scatteraq}
G.~B.~Gelmini, P.~Gondolo and E.~Roulet,
Nucl.\ Phys.\ B {\bf 351}, 623 (1991);
M.~Srednicki and R.~Watkins,
Phys.\ Lett.\ B {\bf 225}, 140 (1989);
M.~Drees and M.~Nojiri,
Phys.\ Rev.\ D {\bf 48}, 3483 (1993)
[arXiv:hep-ph/9307208];
M.~Drees and M.~M.~Nojiri,
Phys.\ Rev.\ D {\bf 47}, 4226 (1993)
[arXiv:hep-ph/9210272];
J.~R.~Ellis, A.~Ferstl and K.~A.~Olive, 
Phys.~Lett.~B  481, (2000) 304,
[arXiv:hep-ph/0001005].


\bibitem{tait}
  G.~Servant and T.~M.~P.~Tait,
  New J.\ Phys.\  {\bf 4}, 99 (2002)
  [arXiv:hep-ph/0209262].

\bibitem{lhdark}
  A.~Birkedal, A.~Noble, M.~Perelstein and A.~Spray,
  Phys.\ Rev.\ D {\bf 74}, 035002 (2006)
  [arXiv:hep-ph/0603077];
  J.~Hubisz and P.~Meade,
  Phys.\ Rev.\ D {\bf 71}, 035016 (2005)
  [arXiv:hep-ph/0411264].


\bibitem{carena}
  M.~Carena, D.~Hooper and P.~Skands,
  Phys.\ Rev.\ Lett.\  {\bf 97}, 051801 (2006)
  [arXiv:hep-ph/0603180];
  M.~Carena, D.~Hooper and A.~Vallinotto,
  arXiv:hep-ph/0611065.



\bibitem{wmap}
  D.~N.~Spergel {\it et al.},
  arXiv:astro-ph/0603449.

\bibitem{uedrelic}
  F.~Burnell and G.~D.~Kribs,
  Phys.\ Rev.\ D {\bf 73}, 015001 (2006)
  [arXiv:hep-ph/0509118];
  K.~Kong and K.~T.~Matchev,
  JHEP {\bf 0601}, 038 (2006)
  [arXiv:hep-ph/0509119];
  M.~Kakizaki, S.~Matsumoto and M.~Senami,
  Phys.\ Rev.\ D {\bf 74}, 023504 (2006)
  [arXiv:hep-ph/0605280].

\bibitem{ewpo}
T.~Appelquist, H.~C.~Cheng and B.~A.~Dobrescu,
Phys.\ Rev.\ D {\bf 64}, 035002 (2001)
[arXiv:hep-ph/0012100];
  I.~Gogoladze and C.~Macesanu,
  Phys.\ Rev.\ D {\bf 74}, 093012 (2006)
  [arXiv:hep-ph/0605207];
  T.~Flacke, D.~Hooper and J.~March-Russell,
  Phys.\ Rev.\ D {\bf 73}, 095002 (2006)
 [Erratum-ibid.\ D {\bf 74}, 019902 (2006)]
  [arXiv:hep-ph/0509352].




\bibitem{capture}
A.~Gould, Astrophys.\ J.\ {\bf 388}, 338 (1991); 
G.~Jungman, M.~Kamionkowski and K.~Griest,
Phys.\ Rept.\  {\bf 267}, 195 (1996)
[arXiv:hep-ph/9506380].


\bibitem{equ}
K.~Griest and D.~Seckel,
Nucl.\ Phys.\ B {\bf 283}, 681 (1987)
[Erratum-ibid.\ B {\bf 296}, 1034 (1988)];
A.~Gould,
Astrophys.\ J.\  {\bf 321}, 571 (1987).

\bibitem{halzen}
  F.~Halzen and D.~Hooper,
  Phys.\ Rev.\ D {\bf 73}, 123507 (2006)
  [arXiv:hep-ph/0510048].


\bibitem{spinfraction}
G.~K.~Mallot,
in {\it Proc. of the 19th Intl. Symp. on Photon and Lepton Interactions at High Energy LP99 } ed. J.A. Jaros and M.E. Peskin,
Int.\ J.\ Mod.\ Phys.\ A {\bf 15S1}, 521 (2000)
[eConf {\bf C990809}, 521 (2000)]
[arXiv:hep-ex/9912040];
W.~M.~Alberico, S.~M.~Bilenky and C.~Maieron,
Phys.\ Rept.\  {\bf 358}, 227 (2002)
[arXiv:hep-ph/0102269].



\bibitem{Hooper:2002gs}
  D.~Hooper and G.~D.~Kribs,
  Phys.\ Rev.\ D {\bf 67}, 055003 (2003)
  [arXiv:hep-ph/0208261].

\bibitem{icecube}
T.~DeYoung  [IceCube Collaboration],
  Int.\ J.\ Mod.\ Phys.\ A {\bf 20}, 3160 (2005);
J.~Ahrens {\it et al.}  [The IceCube Collaboration],
  Nucl.\ Phys.\ Proc.\ Suppl.\  {\bf 118}, 388 (2003)
  [arXiv:astro-ph/0209556].

\bibitem{km3}
  P.~Sapienza,
  Nucl.\ Phys.\ Proc.\ Suppl.\  {\bf 145}, 331 (2005).

\bibitem{lhneu}
  M.~Perelstein and A.~Spray,
  arXiv:hep-ph/0610357.


\bibitem{pamela}
  A.~Morselli and P.~Picozza,
{\it Prepared for 4th International Workshop on the Identification of Dark Matter (IDM 2002), York, England, 2-6 Sep 2002}.



\bibitem{ams02}
  M.~Sapinski  [AMS Collaboration],
  Acta Phys.\ Polon.\ B {\bf 37}, 1991 (2006);
  C.~Goy  [AMS Collaboration],
  J.\ Phys.\ Conf.\ Ser.\  {\bf 39}, 185 (2006).


\bibitem{baltzpos}
E.~A.~Baltz and J.~Edsjo,
Phys.\ Rev.\ D {\bf 59} (1999) 023511
[arXiv:astro-ph/9808243].

\bibitem{kkpos}
  D.~Hooper and G.~D.~Kribs,
  Phys.\ Rev.\ D {\bf 70}, 115004 (2004)
  [arXiv:hep-ph/0406026].



\bibitem{heat}
S.~W.~Barwick {\it et al.}  [HEAT Collaboration],
Astrophys.\ J.\  {\bf 482}, L191 (1997)
[arXiv:astro-ph/9703192];
S.~Coutu {\it et al.}  [HEAT-pbar Collaboration],
in Proceedings of 27th ICRC (2001).

\bibitem{ams01}
Olzem, Jan [AMS Collaboration],
Talk given at the 7th UCLA Symposium on Sources and Detection of Dark Matter and Dark Energy in the Universe, Marina del Ray, CA, Feb 22-24, 2006.


\bibitem{lhpos}
  M.~Asano, S.~Matsumoto, N.~Okada and Y.~Okada,
  arXiv:hep-ph/0602157.

\bibitem{diffusion}
W.~R.~Webber, M.~A.~Lee and M.~Gupta,
Astrophys.\ J.{\bf 390} (1992) 96;
I.~V.~Moskalenko, A.~W.~Strong, S.~G.~Mashnik and J.~F.~Ormes,
Astrophys.\ J.\  {\bf 586}, 1050 (2003)
[arXiv:astro-ph/0210480];
I.~V.~Moskalenko and A.~W.~Strong,
Phys.\ Rev.\ D {\bf 60}, 063003 (1999)
[arXiv:astro-ph/9905283].


\bibitem{btoc}
D.~Maurin, F.~Donato, R.~Taillet and P.~Salati,
Astrophys.\ J.\  {\bf 555}, 585 (2001)
[arXiv:astro-ph/0101231];
D.~Maurin, R.~Taillet and F.~Donato,
Astron.\ Astrophys.\  {\bf 394}, 1039 (2002)
[arXiv:astro-ph/0206286].




\bibitem{silkpos}
  D.~Hooper and J.~Silk,
  Phys.\ Rev.\ D {\bf 71}, 083503 (2005)
  [arXiv:hep-ph/0409104].

\bibitem{secbg}
  I.~V.~Moskalenko and A.~W.~Strong,
  Astrophys.\ J.\  {\bf 493}, 694 (1998)
  [arXiv:astro-ph/9710124].









\bibitem{gc}
  L.~Bergstrom, P.~Ullio and J.~H.~Buckley,
  Astropart.\ Phys.\  {\bf 9}, 137 (1998)
  [arXiv:astro-ph/9712318];
  D.~Hooper and B.~L.~Dingus,
  Phys.\ Rev.\ D {\bf 70}, 113007 (2004)
  [arXiv:astro-ph/0210617].



\bibitem{Bergstrom:2005qk}
  L.~Bergstrom and D.~Hooper,
  Phys.\ Rev.\ D {\bf 73}, 063510 (2006)
  [arXiv:hep-ph/0512317].




\bibitem{dwarfs}
  N.~W.~Evans, F.~Ferrer and S.~Sarkar,
  Phys.\ Rev.\ D {\bf 69}, 123501 (2004)
  [arXiv:astro-ph/0311145];
  S.~Profumo and M.~Kamionkowski,
  JCAP {\bf 0603}, 003 (2006)
  [arXiv:astro-ph/0601249].



\bibitem{glast}
  N.~Gehrels and P.~Michelson,
  Astropart.\ Phys.\  {\bf 11}, 277 (1999);
  S.~Peirani, R.~Mohayaee and J.~A.~de Freitas Pacheco,
  Phys.\ Rev.\ D {\bf 70}, 043503 (2004)
  [arXiv:astro-ph/0401378].

\bibitem{hess}
F.~Aharonian {\it et al.}  [The HESS Collaboration],
  Astron.\ Astrophys.\  {\bf 425}, L13 (2004)
  [arXiv:astro-ph/0408145];
For example, see F.~Ahronian, Talk at TeV Particle Astrophysics Workshop, Batavia, USA, July 2005. 


\bibitem{magic}
  J.~Albert {\it et al.}  [MAGIC Collaboration],
  Astrophys.\ J.\  {\bf 638}, L101 (2006)
  [arXiv:astro-ph/0512469].




\bibitem{veritas}
  H.~Krawczynski, D.~A.~Carter-Lewis, C.~Duke, J.~Holder, G.~Maier, S.~Le Bohec and G.~Sembroski,
  Astropart.\ Phys.\  {\bf 25}, 380 (2006)
  [arXiv:astro-ph/0604508].




\bibitem{Bergstrom:2004cy}
  L.~Bergstrom, T.~Bringmann, M.~Eriksson and M.~Gustafsson,
  Phys.\ Rev.\ Lett.\  {\bf 94}, 131301 (2005)
  [arXiv:astro-ph/0410359].

\bibitem{Birkedal:2005ep}
  A.~Birkedal, K.~T.~Matchev, M.~Perelstein and A.~Spray,
  arXiv:hep-ph/0507194.


\bibitem{lines}
  L.~Bergstrom and P.~Ullio,
  Nucl.\ Phys.\ B {\bf 504}, 27 (1997)
  [arXiv:hep-ph/9706232];
  P.~Ullio and L.~Bergstrom,
  Phys.\ Rev.\ D {\bf 57}, 1962 (1998)
  [arXiv:hep-ph/9707333].



\bibitem{andy}
  D.~Hooper and A.~M.~Taylor,
  arXiv:hep-ph/0607086.


\bibitem{malcolm}
  L.~Bergstrom, M.~Fairbairn and L.~Pieri,
  arXiv:astro-ph/0607327.

\bibitem{compression}
  F.~Prada, A.~Klypin, J.~Flix, M.~Martinez and E.~Simonneau,
  arXiv:astro-ph/0401512.

\bibitem{spike}
  P.~Ullio, H.~Zhao and M.~Kamionkowski,
  Phys.\ Rev.\ D {\bf 64}, 043504 (2001)
  [arXiv:astro-ph/0101481];
  G.~Bertone, G.~Sigl and J.~Silk,
  Mon.\ Not.\ Roy.\ Astron.\ Soc.\  {\bf 337}, 98 (2002)
  [arXiv:astro-ph/0203488];
  P.~Gondolo and J.~Silk,
  Phys.\ Rev.\ Lett.\  {\bf 83}, 1719 (1999)
  [arXiv:astro-ph/9906391].



\bibitem{nfw}
 J.~F.~Navarro, C.~S.~Frenk and S.~D.~M.~White,
  Astrophys.\ J.\  {\bf 462}, 563 (1996)
  [arXiv:astro-ph/9508025];
  J.~F.~Navarro, C.~S.~Frenk and S.~D.~M.~White,
  Astrophys.\ J.\  {\bf 490}, 493 (1997).



\bibitem{moore}
  B.~Moore, S.~Ghigna, F.~Governato, G.~Lake, T.~Quinn, J.~Stadel and P.~Tozzi,
  Astrophys.\ J.\  {\bf 524}, L19 (1999).


\bibitem{gabi}
  G.~Zaharijas and D.~Hooper,
  %
  Phys.\ Rev.\ D {\bf 73}, 103501 (2006)
  [arXiv:astro-ph/0603540].













\end{thebibliography}
\end{document}